\documentclass[lettersize,journal]{IEEEtran}
\usepackage[ruled,linesnumbered]{algorithm2e}

\usepackage{ulem}
\usepackage{multirow}
\usepackage[T1]{fontenc} 
\usepackage{bm} 
\usepackage{amsmath}
\usepackage{amssymb}
\usepackage{graphicx}
\usepackage{subfigure}
\usepackage{cite}
\usepackage{soul}
\usepackage{color}
\usepackage{stfloats} 
\usepackage{multirow}
\usepackage{url}
\usepackage{xurl}

\ifCLASSINFOpdf
\else
\fi

\begin{document}
	
	\title{A Cost-Driven Fuzzy Scheduling Strategy for Intelligent Workflow Decision Making Systems in Uncertain Edge-Cloud Environments}
	%
	%
	%
	
	\author{Bing Lin, Chaowei Lin, and Xing Chen, \textit{Member}, \textit{IEEE}
		\thanks{			
			This work is partly supported by the Natural Science Foundation of China under Grant No. 62072108, the Natural Science Foundation of Fujian Province for Distinguished Young Scholar No. 2020J06014, and the University-Industry Cooperation of Fujian Province under Grant No. 2022H6024. 
			
			Bing Lin is with the College of Physics and Energy, Fujian Normal University, Fujian Provincial Key Laboratory of Quantum Manipulation and New Energy Materials, Fuzhou, 350117, China,  Fujian Provincial Collaborative Innovation Center for Advanced High-Field Superconducting Materials and Engineering, Fuzhou, 350117, China, and  School of Computer Science, Peking University, 100871, China.
			E-mail: WheelLX@163.com.
			
			Chaowei Lin and Xing Chen are with the College of Mathematics and Computer Science, Fuzhou University, Fuzhou, 350118, China, and with Fujian Provincial Key Laboratory of Network Computing and Intelligent Information Processing, Fuzhou, 350118, China. E-mail: cwlin1998@foxmail.com, chenxing@fzu.edu.cn.

%
			
			
			
			
%

		}
	}

	\markboth{}%
	{Shell \MakeLowercase{\textit{et al.}}: A Sample Article Using IEEEtran.cls for IEEE Journals}
	%



	\maketitle
	
	\begin{abstract}
		Workflow decision making is critical to performing many practical workflow applications. Scheduling in edge-cloud environments can address the high complexity problem of workflow applications, while decreasing the data transmission delay between the cloud and end devices. However, because of the heterogeneous resources in edge-cloud environments and the complicated data dependencies among the tasks in a workflow, significant challenges for workflow scheduling remain, including the selection of an optimal tasks-servers solution from the possible numerous combinations. The existing studies have been mainly done subject to rigorous conditions without fluctuations, ignoring the fact that workflow scheduling is typically present in uncertain environments. In this study, we focus on reducing the  execution cost of workflow applications mainly caused by task computation and data transmission, while satisfying the workflow deadline in uncertain edge-cloud environments. The Triangular Fuzzy Numbers (TFNs) are adopted to represent the task processing time and data transferring time. A cost-driven fuzzy scheduling strategy based on an Adaptive Discrete Particle Swarm Optimization (ADPSO) algorithm  is proposed, which employs the operators of Genetic Algorithm (GA). This strategy introduces the randomly two-point crossover operator, neighborhood mutation operator, and adaptive multipoint mutation operator of GA to effectively avoid converging on local optima. The experimental results show that our strategy can effectively reduce the workflow execution cost in uncertain edge-cloud environments, compared with other benchmark solutions.
				
	\end{abstract}
	
	\begin{IEEEkeywords}
		Decision-making, Uncertain edge-cloud environments, Workflow applications, Cost-driven scheduling strategy,  Deadline constraints
	\end{IEEEkeywords}

	%
	\IEEEpeerreviewmaketitle

	\section{Introduction}
	%
	%
	%
	%
	\IEEEPARstart{W}{orkflows} are widely exploited for the modeling of complicated applications, such as DNN-based applications \cite{9294146,8941306}. 
	Such a workflow is usually  computing-intensive, typically composed of tens of interdependent tasks. Indeed, workflow scheduling is essential as its result could directly affect the performance of workflow applications \cite{HAN2020101837}. Due to the complicated structure of a workflow and the sophisticated data dependencies between the tasks in a workflow, completing workflow scheduling in time can be rather challenging even with the use of a high-performance computing platform. 
	
	To tackle the aforementioned challenge, some studies have been focused on workflow scheduling in the cloud computing environment \cite{7508444,ISMAYILOV2020307,9095379}. Such work is mainly oriented towards cloud service providers  (e.g., Amazon EC2, Rackspace, and GoGrid) which provide virtual resources to the end customers \cite{SHU202112,7416240}. The underlying techniques have been developed in an attempt to rationally schedule the dependent tasks among the virtual resources through the pay-as-you-go method. However, they are generally aimed at reducing the workflow completion time and improving the resource utilization, with far less a focus on optimizing the  execution cost of workflow applications. They mostly ignore the performance variation between virtual machines with different configurations. In terms of scheduling workflow in the cloud, it might
	increase the traffic load of core networks and cause high latency due to massive data transmission between clients and the cloud \cite{8668458}.
	
	Edge computing provides an essential technology to improve performance in workflow scheduling \cite{7931566}. It enhances the computing ability of a mobile network through deploying the computation and storage resources around the edges of the mobile web, thereby providing the service with high broadband and low delay to the clients. Workflow scheduling in edge-cloud environments not only meets the compute-intensive requirements of workflow applications, but also effectively reduces data transmission delays, while scheduling the data-intensive tasks to the edge and the compute-intensive tasks to the cloud \cite{9519636}.
	Nonetheless, the service nodes (namely, the virtual machines in the cloud and the servers in the edge) are generally heterogeneous and their processing capacities can be rather different, as well as the cost-performance in terms of load/energy consumption \cite{9497712}. To reduce the workflow execution cost while satisfying deadline constraints, significant difficulties remain in rationally scheduling the data-dependent tasks for efficient data transmissions and task executions.

	The existing studies on workflow scheduling are mainly carried out subject to certain conditions (e.g., assuming that the performance of the service nodes, bandwidth, and other factors are steady, without fluctuation) \cite{8561182,8360973,9298863}. However, in the practical scheduling processes, the CPUs of the service nodes and bandwidths between them always fluctuate, which may have considerable impact upon workflow scheduling. Whilst uncertainties in scheduling have been addressed, the relevant work mainly focused on Fuzzy Job Shop Scheduling Problem (FJSSP) \cite{8626515,9120283} or task scheduling for the real-time embedded systems \cite{8967002}. 
	
	The above methods have good inspirations for the workflow scheduling in uncertain edge-cloud environments. However, they suffer from the following limitations:	
	
	\begin{itemize}

		\item Due to the uncertain conditions, the task processing time and data transferring time are usually difficult to determine. Hence, existing workflow scheduling strategies can not be directly applied to the uncertain edge-cloud environments. 
		
		\item Due to the heterogeneity among servers, there is a non-negligible difference in their execution performance. Hence, the existing methods are difficult to select an optimal cost-driven tasks-servers scheduling solution from the numerous servers.

		\item Existing workflow scheduling strategies mainly considers factors such as the server load balancing and energy consumption, the workflow completion time and execution cost, and less consideration is given to the comprehensive cost optimization of task computing and data transmission within deadline constraints.
			
	\end{itemize}
		
	To address the above questions, we develop a novel workflow scheduling strategy to reduce the workflow execution cost caused by task computation and data transmission, while satisfying the required deadline constraints in uncertain edge-cloud environments.	
	The major contributions of this work are summarized below:	
		
	\begin{itemize}

		\item To reasonably model workflow scheduling in uncertain edge-cloud environments, Triangular Fuzzy Numbers (TFNs) \cite{7782438} are adopted to represent the task processing time and data transferring time. 
		
		\item To reduce the comprehensive cost of task computing and data transmission within deadline constraints, a cost-driven workflow scheduling strategy based on an Adaptive Discrete Particle Swarm Optimization (ADPSO) algorithm employing the operators of Genetic Algorithm (GA) is proposed, which improves the exploration and exploitation of scheduling strategies to obtain a better result.
		
		\item The extensive simulation experiments are conducted. The performance results demonstrate that the proposed strategy can achieve the superior performance than other classic methods with respect to commonly adopted benchmark datasets.	
	\end{itemize}

	The rest of this paper is organized as follows: Section II briefly reviews the related work. Section III presents the problem of workflow scheduling in uncertain edge-cloud environments. Section IV describes our proposed workflow scheduling strategy in detail. Section V analyzes the performance of our strategy through experimental studies in comparison with the state-of-the-art scheduling strategies. Finally, Section VI summarizes the work and outlines relevant future research directions.

	\section{Related Work}
	
	A workflow model, used to simulate and analyze the workflow applications in the real world, consists of a set of the computational tasks linked through control and data dependencies \cite{7145406}. Workflow scheduling is essential as its result could directly affect the performance of workflow applications.
	
	Many research efforts have been launched to workflow scheduling in cloud computing. Yuan \textit{et al}. \cite{7508444} considered the cost minimization of data centers in private cloud. They proposed a Temporary Task Scheduling Algorithm (TTSA) that could efficiently schedule all arriving tasks to the private or public clouds. 
	This method effectively reduces the cost of the private cloud while satisfying all tasks' delay constraints. 
	Meng \textit{et al}. \cite{9095379} proposed a security-aware scheduling method based on the PSO algorithm for real-time resource allocation across heterogeneous clouds. Experimental results showed that this strategy could achieve a good balance between scheduling and security performance. Pham \textit{et al}. \cite{9090348} considered the fulfilment and interruption rates of the volatile resources in order to reflect the instability of the cloud infrastructure. In that work, a novel evolutionary multi-objective workflow scheduling approach was proposed for generating a set of trade-off solutions, whose makespan and cost were superior to the state-of-the-art algorithms. 
	Paknejad \textit{et al.} \cite{PAKNEJAD202112} proposed an enhanced multi-objective co-evolutionary algorithm, called ch-PICEA-g, for workflow scheduling in cloud environment. Experiments results indicated that the proposed algorithm outperformed its counterparts in terms of different performance metrics, such as cost, makespan, and energy consumption.
	It is of practical significance for scheduling workflow in cloud computing. However, it might increase the traffic load of
	core networks and cause high latency due to massive data
	transmission between clients and the cloud.
	
	Edge computing can effectively reduce the system delay of workflow scheduling \cite{9519636,7931566,JARARWEH202042}.  Workflow scheduling in edge-cloud environments has recently drawn great interest. For instance,
	Xie \textit{et al}. \cite{XIE2019361} designed a novel Directional and Non-local-Convergent PSO (DNCPSO) algorithm to simultaneously optimize the completion time and execution cost of the workflow. Experimental results demonstrated that DNCPSO could achieve better performance than other classic algorithms. 	
	Peng \textit{et al}. \cite{8743291} proposed a node reliability model to evaluate resource reliability in Mobile Edge Computing (MEC) environments, defining workflow scheduling as an optimization problem and solving it by an algorithm based on Krill-Herd \cite{GANDOMI20124831}. 
	Through experiments based on real workflow applications and mobile user contract tracking, it had proven that the performance of this method was significantly better than the traditional methods in terms of success rate and makespan.
	However, existing research on workflow scheduling in edge-cloud environments hardly considers the comprehensive cost optimization for task computation and data transmission.
	
	In real-world practice, the performance of service nodes and the bandwidth may fluctuate while scheduling a workflow application. Initial work exists and deals with scheduling in uncertain computing environments, but such work is mainly oriented towards intelligent manufacturing systems. In particular, Lei \cite{LEI2010610} represented the fuzzy processing time and fuzzy due-date with TFNs and trapezoidal fuzzy numbers, respectively, while introducing an improved fuzzy max operation to investigate the FJSSP. In that work, so-called availability constraints are employed for maximizing the satisfaction level of customers. Sun \textit{et al} \cite{8626515} also used TFNs to describe the processing time to cope with the FJSSP  problem, where an effective hybrid Cooperative Evolution Algorithm (hCEA) was proposed for minimizing the fuzzy makespan. Fortemps \cite{649907} expressed an uncertain duration as a six-point fuzzy number, thereby establishing a fuzzy scheduling model to minimize the fuzzy completion time for job shop scheduling problem. Similarly, Li \textit{et al} \cite{LI2019105585} used TFNs to capture the uncertainty of fuzzy processing time and introduced a uniform parallel machine scheduling with such processing time representation under fuzzy resource consumption constraints, minimizing the makespan. 
	
	Despite the aforementioned remarkable developments in the relevant research area, an important open issue remains on fuzzy workflow scheduling that is of great practical significance: Workflow scheduling that considers the fuzzy task processing time and fuzzy data transferring time in uncertain edge-cloud environments. Inspired by this observation, the reminder of this paper will establish a novel approach to cost-driven scheduling for deadline-based workflows in uncertain edge-cloud environments.

	\section{System Model and Definitions}
	
	In this section, the workflow scheduling in certain environments is described firstly. Then we further elaborate the workflow scheduling in uncertain environments. Thirdly, the operations for TFNs in fuzzy workflow scheduling are introduced in detail. Finally, an example of cost-driven scheduling for a deadline-based workflow application in uncertain edge-cloud environments is illustrated.   
	
	\subsection{Workflow Scheduling in Certain Environments}
	
	The workflow scheduling framework proposed in this study consists of three main components, \textit{i.e.}, the edge-cloud environments, a deadline-based workflow, and a cost-driven scheduler.
	
	A certain environment means that there is no fluctuation during workflow scheduling and execution. The edge-cloud environments $\bm{S} = \left\{ {{\bm{S}_{cloud}},{\bm{S}_{edge}}} \right\}$ consist of the cloud and edge, where there are different computing nodes (\textit{i.e.},  virtual machines in the cloud and severs in the edge). For simplicity, we use `servers' to denote the computing nodes in the cloud and edge with a uniform representation. There are $n$ servers in the cloud ${\bm{S}_{cloud}} = \left\{ {{s_1},{s_2}, \ldots ,{s_n}} \right\}$, and $m$ servers in the edge ${\bm{S}_{edge}} = \left\{ {{s_{n + 1}},{s_{n + 2}}, \ldots ,{s_{n + m}}} \right\}$. A server $s_i$ is denoted by Eq. (\ref{eq1}).
	
	\begin{equation}\label{eq1}
		{s_i} = \left( {{\zeta}_i,{\varpi}_i ,{p_i},c_i^{com},{\lambda _i},{f_i}} \right),
	\end{equation}

	\noindent where ${\zeta}_i$ and ${\varpi}_i$ are the booting time and shutdown time of the server $s_i$, respectively;  ${p_i}$ is
	the processing capacity of the server $s_i$; $c_i^{com}$ is the computation cost per time unit ${\lambda _i}$, which is a specific time unit for the server $s_i$; ${f_i}=\{ 0,1\} $ refers to the platform to which the server $s_i$ belongs. Note that when ${f_i}=0$, $s_i$ belongs to the cloud with powerful processing capacity. Otherwise, $s_i$ belongs to the edge with normal processing capacity.
	
	The bandwidth $b_{i,j}$ between any two different servers is denoted by Eq. (\ref{eq2}).
	
	\begin{equation}\label{eq2}
		{b_{i,j}} = \left( {{\beta _{i,j}},c_{i,j}^{tran}} \right),
	\end{equation}
	
	\noindent where ${\beta _{i,j}}$ is the value of the bandwidth $b_{i,j}$, 
	$\forall i,j=1,2,...,|\bm{S}| ,  i \ne j$; 	$c_{i,j}^{tran}$ is the data transmission cost per GB from the server $s_i$ to $s_j$. 
		
	A workflow can be described as a directed acyclic graph (DAG) $\bm{W} = \left( {\bm{V},\bm{E}} \right)$, where 
	$\bm{V} = \left\{ {{v_1},{v_2}, \ldots ,{v_l}} \right\}$ is a finite set of $l$ tasks, and 
	$\bm{E} = \{ {e_{i,j}} = \left. {\left\langle {{v_i},{v_j}} \right\rangle } \right|{v_i},{v_j} \in \bm{V}, \forall i \ne j \} $ is a finite set of directed arcs.
	Each directed arc ${e_{i,j}={\left\langle {{v_i},{v_j}} \right\rangle }}$ indicates that there is a dataset $d_{i,j}$ transferred from the task $v_i$ to $v_j$, and  $v_j$ cannot be executed until $v_i$ is finished. For an arc ${e_{i,j}={\left\langle {{v_i},{v_j}} \right\rangle }}$, $v_i$ is called the immediate predecessor task of $v_j$, and $v_j$ is called the immediate  successor task of $v_i$.	
	In addition, a workflow has a corresponding deadline constraint $D\left(\bm{W} \right)$. When a  workflow is completed within its deadline based on a specific scheduling strategy, this strategy is called as a feasible solution. 
	
			
	Suppose that the processing time of the task $v_i$ on the server $s_j$ is described as $t_{com}(v_i,s_j)$. Owing to its popularity, the serial
	processing model \cite{7364271} is adopted herein. It expresses that a task is processed on only
	one server, and a server can  process only one task concurrently.
	The data transferring time $t_{tran}(d_{i,j},s_k,s_l)$ is denoted by Eq. (\ref{eq3}).
	
	\begin{equation}\label{eq3}
	{t_{tran}}\left( {{d_{i,j}},{s_k},{s_l}} \right) = \frac{{{d_{i,j}}}}{{{\beta _{k,l}}}},
	\end{equation}
	
	\noindent where $t_{tran}(d_{i,j},s_k,s_l)$ is the time to transfer the dataset $d_{i,j}$ from the server $s_k$ to $s_l$. If $s_k$ and $s_l$ are the same server, the data transferring time is 0. 

	A cost-driven scheduler aims to reduce the workflow execution cost mainly caused by task computation and data transmission, while satisfying its deadline in edge-cloud environments. A scheduling strategy $\Psi$ is denoted by Eq. (\ref{eq4}).
	
	\begin{equation}\label{eq4}
	\Psi  = \left( {\bm{W},\bm{S},\bm{M},{t_{t}},{c_{t}}} \right),
	\end{equation}
	
	\noindent where $ \bm{M} = \{ \left( {{v_i},{s_j}} \right) \cup \left( {{d_{k,l}},{s_r},{s_t}} \right)|{v_i} \in \bm{V},{d_{k,l}} \in \bm{E}, {s_j}, $ ${s_r}, {s_t} \in \bm{S} \}$ is the mapping from the tasks and datasets to the servers; $t_{t}$ is the workflow completion time, and $c_{t}$ is the workflow execution cost in edge-cloud environments with a given scheduling strategy.
	
	There are two subsets in the mapping $\bm{M}$: $\left( {{v_i},{s_j}} \right)$ indicates that the task $v_i$ is executed on the server $s_j$, and $\left( {{d_{k,l}},{s_r},{s_t}} \right)$ implies that the dataset $d_{k,l}$ is transferred from the server $s_r$ to $s_t$. When the subset ${\bm{M}_{\bm{V}}} = \left\{ {\left( {{v_i},{s_j}} \right)|{v_i} \in \bm{V},{s_j} \in \bm{S}} \right\}$ is determined, the other subset 
	${\bm{M}_{\bm{E}}} = \left\{ {\left( {{d_{k,l}},{s_r},{s_t}} \right)|{d_{k,l}} \in \bm{E},{s_r},{s_t} \in \bm{S}} \right\}$ will be determined. Therefore, the mapping $\bm{M}$ is equivalent to ${\bm{M}_{\bm{V}}}$ as Eq. (\ref{eq5}).
	
	\begin{equation}\label{eq5}
	{\bm{M}} = {{\bm{M}}_{\bm{V}}} = \left\{ {\left( {{v_i},{s_j}} \right)|{v_i} \in {\bm{V}},{s_j} \in {\bm{S}}} \right\}.
	\end{equation}
	
	When the mapping $\bm{M}$ is determined, the servers processing all tasks are determined with a specific scheduling strategy $\Psi$. Due to the data dependencies between the tasks in a workflow, the execution order of each task is relatively fixed. Each task $v_i$ will have the start time ${t_{start}}\left( {{v_i}} \right)$ and the end time ${t_{end}}\left( {{v_i}} \right)$ once the corresponding $\bm{M}$ is determined. The workflow completion time can be denoted by Eq. (\ref{eq6}).

	\begin{equation}\label{eq6}
	{t_{t}} = \mathop {\max }\limits_{{v_i} \in \bm{V}} \left\{ {{t_{end}}\left( {{v_i}} \right)} \right\}.
	\end{equation}
	
	We assume that the shutdown time ${\varpi}_i$ of $s_i$ is equal to the end time of the last task on it. The total execution cost ${c_{t}}$ of scheduling a workflow in edge-cloud environments is determined by Eqs. (\ref{eq7}-\ref{eq9}).
	
	\begin{gather}
	{c_{t}} = {c_{com}} + {c_{tran}}, \label{eq7}\\
	{c_{com}} = \mathop \sum \limits_{i = 1}^{\left| \bm{S} \right|} c_i^{com} \cdot \left\lceil {\frac{{\varpi}_i-{\zeta}_i}{{{\lambda _i}}}} \right\rceil , \label{eq8}\\
	{c_{tran}} = \mathop \sum \limits_{{v_j} \in {\bm{V}}} \mathop \sum \limits_{{v_k} \in {\bm{V}}} c_{r,t}^{tran} \cdot {d_{j,k}},\left( {{v_j},{s_r}} \right),\left( {{v_k},{s_t}} \right) \in {\bm{M}}, \label{eq9}
	\end{gather}
	
	\noindent where $c_{com}$ is the task computation cost, and $c_{tran}$ is the data transmission cost.
	
	In summary, the scheduling strategy for a workflow in certain environments can be described by Eq. (\ref{eq10}), which indicates that the scheduler pursues to minimize the total  workflow execution cost $c_{t}$, while satisfying its deadline $D(\bm{W})$.
		
	\begin{equation}\label{eq10}
	\left\{ {\begin{array}{*{20}{l}}
		{\min {\kern 1pt} {\kern 1pt} {\kern 1pt} {\kern 1pt} {c_{t}};} \vspace{1ex}\\
		{s.t.{\kern 1pt} {\kern 1pt} {\kern 1pt} {\kern 1pt} {t_{t}} \le D\left( {\bm{W}} \right).}
		\end{array}} \right.
	\end{equation}	
	
	\subsection{Workflow Scheduling in Uncertain Environments} \label{sec3.2}
	
	Uncertain environments mean that there are fluctuations during workflow scheduling and execution. The task processing time and data transferring time are uncertain due to the fluctuations of server processing capacity and bandwidth, respectively. Fuzzy set  are employed to reflect the uncertainties during workflow execution,  while the task processing time and data transferring time are represented as TFNs.
	
\begin{figure}[htbp]
		\centering
		\includegraphics[scale=0.7]{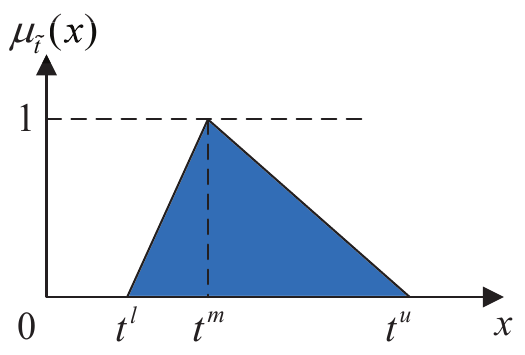}
		\caption{The membership function ${\mu _{\tilde t}}\left( x \right)$ of a TFN $\tilde t$}
		\label{fig1}
	\end{figure}

	The membership function ${\mu _{\tilde t}}\left( x \right)$ of a TFN $\tilde t = \left( {{t^l},{t^m},{t^u}} \right)$ can be denoted by Eq. (\ref{eq11}), which is represented graphically as Fig. \ref{fig1} \cite{9064649}.
	
	\begin{equation}\label{eq11}
	{\mu _{\tilde t}}\left( x \right) = \left\{ {\begin{array}{*{20}{l}}
		{\frac{{x - {t^l}}}{{{t^m} - {t^l}}},x \in \left[ {{t^l},{t^m}} \right]}\vspace{1ex};\\
		{\frac{{x - {t^u}}}{{{t^m} - {t^u}}},x \in \left[ {{t^m},{t^u}} \right]}\vspace{1ex};\\
		{{\kern 1pt} {\kern 1pt} {\kern 1pt} {\kern 1pt} {\kern 1pt} {\kern 1pt} {\kern 1pt} {\kern 1pt} {\kern 1pt} {\kern 1pt} {\kern 1pt} 0 {\kern 1pt} {\kern 1pt} {\kern 1pt} {\kern 1pt} {\kern 1pt} {\kern 1pt} {\kern 1pt} {\kern 1pt} {\kern 1pt} {\kern 1pt} {\kern 1pt} ,x \in \left( { - \infty ,{t^l}} \right) \cup \left( {{t^u}, + \infty } \right)}.
		\end{array}} \right.
	\end{equation}	
	
	\noindent Where ${t^m}$ is the normal (namely, the most possible) value of the fuzzy variable $\tilde t$; ${t^l}$ and ${t^u}$ are the lower and upper limit values of $\tilde t$, respectively. When ${t^l} = {t^m} = {t^u}$, $\tilde t$ is a certain number. A fuzzy variable $\tilde \tau $ in uncertain environments corresponds to a variable $\tau $ in certain environments. According to the principles of fuzzy set theory \cite{ext-principle},  the scheduling strategy for a workflow in uncertain environments can be defined as Eq. (\ref{eq12}).

	\begin{equation}\label{eq12}
	\left\{ {\begin{array}{*{20}{l}}
		{\min {\kern 1pt} {\kern 1pt} {\kern 1pt} {\kern 1pt} {{\tilde c}_{t}}} \vspace{1ex};\\
		{s.t.{\kern 1pt} {\kern 1pt} {\kern 1pt} {\kern 1pt} {{\tilde t}_{t}} \le D\left( {\bm{W}} \right)}.
		\end{array}} \right.
	\end{equation}
	
	\noindent Where ${\tilde c_{t}}$ is the fuzzy total workflow execution cost, and ${\tilde t_{t}}$ is the fuzzy workflow completion time. Both fuzzy variables (\textit{i.e.}, ${\tilde c_{t}}$ and ${\tilde t_{t}}$) are represented by TFNs. 
	For the target $\min {\kern 1pt} {\kern 1pt} {\tilde c_{t}}$, an equivalent representation through its mean value and the standard deviation  can be introduced \cite{LEE1988887}, 
	where the objective function given in Eq. (\ref{eq12}) is equivalent to that Eq. (\ref{eq13}).

	\begin{gather}
	\begin{array}{l}
	\min {\kern 1pt} {\kern 1pt} {\kern 1pt} {\kern 1pt} {{\tilde c}_{t}} = \left( {{c^l},{c^m},{c^u}} \right) \Rightarrow \vspace{1ex}\\
	\min {\kern 1pt} {\kern 1pt} {\kern 1pt} {\kern 1pt} m\left( {{{\tilde c}_{t}}} \right) + \eta  \cdot std\left( {{{\tilde c}_{t}}} \right),\eta  \ge 0. \vspace{1ex}
	\end{array} \label{eq13}
		\end{gather}
	
	\noindent Where $m\left( {{{\tilde c}_{t}}} \right)$ and $std\left( {{{\tilde c}_{t}}} \right)$ are the mean value and standard deviation of ${\tilde c_{t}}$, and $\eta $ is the weighting factor of 
	$std\left( {{{\tilde c}_{t}}} \right)$. According to the work of Lee and Li \cite{LEE1988887},  the mean value and standard deviation of a TFN can be defined through uniform distribution and proportional distribution. Therefore, $m\left( {{{\tilde c}_{t}}} \right)$ and $std\left( {{{\tilde c}_{t}}} \right)$ can be computed as Eqs. (\ref{eq14}) and (\ref{eq15}), respectively.

	\begin{gather}
	m\left( {{{\tilde c}_{t}}} \right) = \frac{{\int {x\tilde c_{t}^2\left( x \right){\rm{d}}x} }}{{\int {\tilde c_{t}^2\left( x \right){\rm{d}}x} }} = \frac{{{c^l} + 2{c^m} + {c^u}}}{4}, \vspace{1ex} \label{eq14}\\
	\begin{array}{rl}
	std\left( {{{\tilde c}_{t}}} \right) &= {\left[ {\frac{{\int {{x^2}\tilde c_{t}^2\left( x \right){\rm{d}}x} }}{{\int {\tilde c_{t}^2\left( x \right){\rm{d}}x} }} - {m^2}\left( {{{\tilde c}_{t}}} \right)} \right]^{1/2}} \vspace{1ex}\\
	&= {\left[ {\frac{{2{{\left( {{c^l} - {c^m}} \right)}^2} + {{\left( {{c^l} - {c^u}} \right)}^2} + 2{{\left( {{c^m} - {c^u}} \right)}^2}}}{{80}}} \right]^{1/2}}.
	\end{array} \label{eq15}
	\end{gather}

	In the process of workflow scheduling, the actual task processing time and data transferring time are more likely to be longer than the estimated values \cite{PALACIOS201574}. A new fuzzification method based on Sun \textit{et al.} \cite{8626515} is proposed to describe the uncertain values (namely, the task processing time and data transferring time). Therefore, the related parameters of a TFN $\tilde t = \left( {{t^l},{t^m},{t^u}} \right)$ are redefined as follows: ${t^m}$ is the estimated time; ${t^l}$ and ${t^u}$ are randomly selected from the interval $\left[ {{\delta _1} \cdot t,t} \right]$ and $\left[ {2t - {t^l},{\delta _2} \cdot t} \right]$, respectively, where  ${\delta _1}$ and ${\delta _2}$ are adjustment coefficients, satisfying that ${\delta _1}<1$,  ${\delta _2}>1$ and ${\delta _2} - 1 > 1 - {\delta _1}$.
	
	Such a TFN will satisfy the constraint ${t^u} - {t^m} \ge {t^m} - {t^l}$. Therefore, the mean value $m\left( {\tilde t} \right)$ of a TFN ${\tilde t}$ becomes that in Eq. (\ref{eq18}), which is more likely to be longer than its estimate. 
	
	\begin{equation}\label{eq18}
	m\left( {\tilde t} \right) = \frac{{{t^l} + 2{t^m} + {t^u}}}{4} \ge {t^m} = t.
	\end{equation}
	
For the constraint ${\tilde t_{t}} \le D\left( {\bm{W}} \right)$, the upper limit value ${t^u}$ of ${\tilde t_{t}}$ should be no more than the deadline constraint in the worst case. Therefore, the constraint given in Eq. (\ref{eq12}) is equivalent to that in Eq. (\ref{eq16}).

	\begin{equation}\label{eq16}
	{\kern 1pt} {\kern 1pt} {\tilde t_{t}} = \left( {{t^l},{t^m},{t^u}} \right) \le D\left( {\bm{W}} \right) \Rightarrow {\kern 1pt} {\kern 1pt} {t^u} \le D\left( {\bm{W}} \right).
	\end{equation}

	In summary, the  scheduling strategy for a workflow in uncertain edge-cloud environments can be formalized as Eq. (\ref{eq17}).

	\begin{equation}\label{eq17}
	\left\{ {\begin{array}{*{20}{l}}
		{\min {\kern 1pt} {\kern 1pt} {\kern 1pt} {\kern 1pt} m\left( {{{\tilde c}_{t}}} \right) + \eta  \cdot std\left( {{{\tilde c}_{t}}} \right)} \vspace{1ex};\\
		{s.t.{\kern 1pt} {\kern 1pt} {\kern 1pt}  {\kern 1pt} {\kern 1pt} {\kern 1pt} {\kern 1pt} {\kern 1pt} {t^u} \le D\left( {\bf{W}} \right)}.
		\end{array}} \right.
	\end{equation}

	\subsection{Operations for TFNs in Fuzzy Workflow Scheduling} \label{sec3.3}

	To construct a feasible schedule for a workflow in uncertain edge-cloud environments, the operations for TFNs (\textit{i.e.}, addition, ranking, max, and multiplication) need to be introduced as follows: 
	
	\subsubsection{Addition Operation}
			
 	Addition operation is used to calculate the end time of tasks. Suppose that the start time and processing time of a task are denoted by $\tilde r = \left( {{r^l},{r^m},{r^u}} \right)$ and $\tilde t = \left( {{t^l},{t^m},{t^u}} \right)$, respectively. The end time of such a task $\tilde e = \left( {{e^l},{e^m},{e^u}} \right)$ is calculated by Eq. (\ref{eq19})  \cite{9064649}.
 	
 	\begin{equation}\label{eq19}
 	\widetilde {\rm{e}} = \tilde r + \tilde t = \left( {{r^l} + {t^l},{r^m} + {t^m},{r^u} + {t^u}} \right).
 	\end{equation}

 	\subsubsection{Ranking Operation}
 	
 	Ranking operation is required to calculate the maximum end time of all the immediate predecessors of the task $v_i$.  	Suppose that the end time of one predecessor and that of another predecessor are $\tilde r = \left( {{r^l},{r^m},{r^u}} \right)$ and $\tilde t = \left( {{t^l},{t^m},{t^u}} \right)$, respectively. The maximum end time of such two predecessors is then calculated with respect to three different situations, following the ranking criterion proposed by Sakawa \textit{et al}. \cite{Sakawa2000Fuzzy}.
 	
 	\begin{itemize}

 	\item If ${\alpha _1}\left( {\tilde r} \right) = \left( {{r^l} + 2{r^m} + {r^u}} \right)/4 > {\alpha _1}\left( {\tilde t} \right) = \left( {{t^l} + 2{t^m} + {t^u}} \right)/4$, then $\tilde r > \tilde t$. \vspace{1ex}
 	
 	\item If ${\alpha _1}\left( {\tilde r} \right) = {\alpha _1}\left( {\tilde t} \right)$ and 
 	${\alpha _2}\left( {\tilde r} \right) = {r^m} > {\alpha _2}\left( {\tilde t} \right) = {t^m}$, then $\tilde r > \tilde t$. \vspace{1ex}
 	
 	\item If ${\alpha _1}\left( {\tilde r} \right) = {\alpha _1}\left( {\tilde t} \right),{\alpha _2}\left( {\tilde r} \right) = {\alpha _3}\left( {\tilde t} \right)$ and ${\alpha _3}\left( {\tilde r} \right) = {r^u} - {r^l} > {\alpha _3}\left( {\tilde t} \right) = {t^u} - {t^l}$, then $\tilde r > \tilde t$.
 	\end{itemize}
 	
 \noindent Note that multiple ranking operations  are recursively performed if there are more than two predecessor tasks.
 	
 	\subsubsection{Max Operation}
 	
 	Max operation is needed to calculate the start time of tasks. Suppose that the maximum end time of all immediate predecessors of the task $v_i$ is $\tilde r = \left( {{r^l},{r^m},{r^u}} \right)$, and the last idle time of a server before processing the task $v_i$ is $\tilde t = \left( {{t^l},{t^m},{t^u}} \right)$. Then, the membership function ${\mu _{\tilde e}}\left( z \right)$ of $v_i$'s start time $\tilde e = \tilde r \vee \tilde t$ is computed by Eq. (\ref{eq20}).
 	  
 	\begin{equation}\label{eq20}
 	\begin{array}{rl}
 	{\mu _{\tilde e}}\left( z \right) &= {\mu _{\tilde r \vee \tilde t}}\left( z \right) = \mathop {\sup }\limits_{z = x \vee y} \min \left( {{\mu _{\tilde r}}\left( x \right),{\mu _{\tilde t}}\left( y \right)} \right)\\
 	& \buildrel \Delta \over = \mathop  \vee \limits_{z = x \vee y} \left( {{\mu _{\tilde r}}\left( x \right) \wedge {\mu _{\tilde t}}\left( y \right)} \right).
 	\end{array}
 	\end{equation}  
 	
\noindent According to the max criterion proposed by Lei \cite{LEI2010610}, the start time of the task $v_i$ can be approximated by Eq. (\ref{eq21}) .
 	
 	\begin{equation}\label{eq21}
 	\tilde e = \tilde r \vee \tilde t \cong \left\{ {\begin{array}{*{20}{c}}
 		{\tilde r,\tilde r \ge \tilde t};\\
 		{\tilde t,\tilde r < \tilde t}.
 		\end{array}} \right.
 	\end{equation} 
 	 	
 	\subsubsection{Multiplication Operation}
 	
 	Multiplication operation is carried out to calculate the 
 	task computation cost and data transmission cost as addressed by Eqs. (\ref{eq8}) and (\ref{eq9}), respectively. The product of a TFN $\tilde t = \left( {{t^l},{t^m},{t^u}} \right)$ and a real number $\kappa $ is computed by Eq. (\ref{eq22})  \cite{9064649}.
 	
 	\begin{equation}\label{eq22}
 	\kappa  \cdot \tilde t = \left( {\kappa \cdot {t^l},\kappa \cdot {t^m},\kappa \cdot {t^u}} \right),\forall \kappa  \in \mathbb{R}.
 	\end{equation}
 	
\noindent Similarly, the quotient of a TFN $\tilde t = \left( {{t^l},{t^m},{t^u}} \right)$ divided by a real number $\upsilon $ can be transformed into the product of such a TFN and another real number as Eq. (\ref{eq23}).  
 	
 	\begin{equation}\label{eq23}
 	\tilde t \div \ell   \buildrel \Delta \over = \kappa  \cdot \tilde t{\rm{\;}},\kappa  = 1/\ell ,\forall \ell  \in \mathbb{R}.
 	\end{equation}
 	
 	\subsection{Illustration of Cost-Driven Workflow Scheduling}
 	
\begin{figure}[htbp]
 		\centering
 		\includegraphics[scale=0.55]{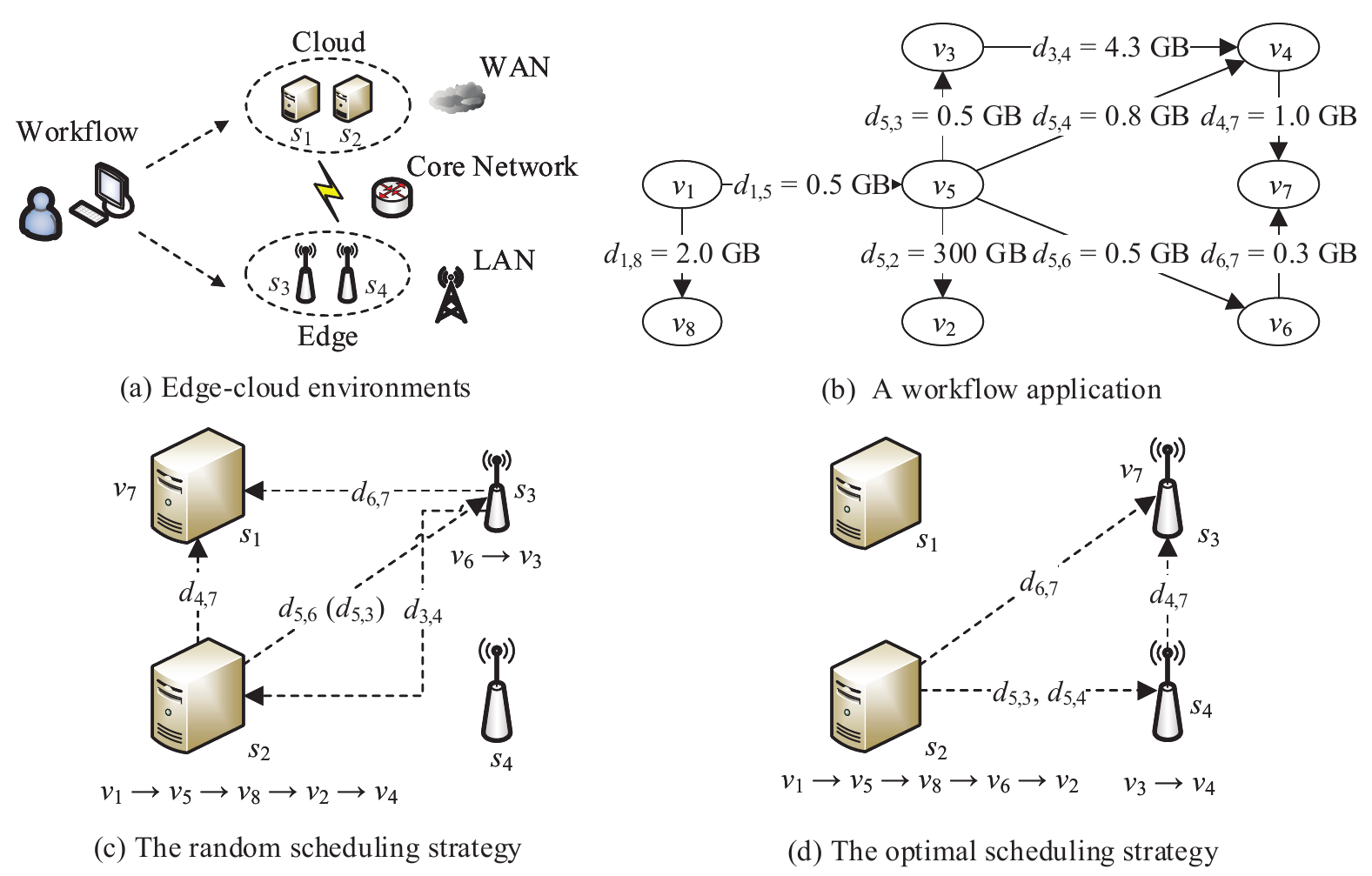} 
 		\caption{An example of cost-driven scheduling for a deadline-based workflow application in uncertain edge-cloud environments.}
 		\label{fig2}
 	\end{figure}

 	Fig. \ref{fig2} presents an example of cost-driven scheduling for a deadline-based workflow in uncertain edge-cloud environments. The edge-cloud environments $\bm{S} = \{ {s_1},{s_2},{s_3},{s_4}\} $  consist of four servers, where ${s_1},{s_2}$ belong to the cloud and ${s_3},{s_4}$ belong to the edge. The workflow application has 8 tasks and 9 datasets, whose deadline is $7.2 \times {10^3}$\textit{s}. The time unit ${\lambda _i}$ is set to 60\textit{s}. Table \ref{tab1} lists the relevant parameters for the bandwidths between different servers. Table \ref{tab2} presents the computation cost per hour for all servers. Table \ref{tab3} shows the fuzzy processing time for each task on the available servers. 
 	
 	\begin{table}[!t]
 		\renewcommand{\arraystretch}{1.3}
 		\caption{Bandwidths between different servers}
 		\label{tab1}
 		\centering
 		\begin{tabular}{ccccc}
 			\hline
 			$f_i$ & $ \leftrightarrow $ & $f_j$ & $\beta_{i,j}($\textit{MB}$\cdot$\textit{s}$^{-1}$) & $c_{i,j}^{tran}($\textit{\$}$\cdot$\textit{GB}$^{-1})$\\
 			\hline
 			0 & $ \leftrightarrow $ & 0 & 2.5 & 0.4 \\
 			0 & $ \leftrightarrow $ & 1 & 1.0 & 0.16 \\
 			1 & $ \leftrightarrow $ & 1 & 12.5 & 0.8 \\
 			\hline
 		\end{tabular}
 	\end{table}
	
 	\begin{table}[!t]
 		\renewcommand{\arraystretch}{1.3}
 		\caption{Computation cost per hour of all servers}
 		\label{tab2}
 		\centering
 		\begin{tabular}{cc}
 			\hline
 			Servers & $c_i^{com}$ (\textit{\$}$\cdot$\textit{h}$^{-1}$ ) \\
 			\hline
 			$s_1$ & 3.5 \\
 			$s_2$ & 13.5 \\
 			$s_3$ & 1.5 \\
 			$s_4$ & 2.0 \\
 			\hline
 		\end{tabular}
 	\end{table}
	
	\begin{table*}[htb]
		\renewcommand{\arraystretch}{1.3}
		\caption{Fuzzy processing time for each task on available servers (unit: \upshape \textit{s})}
		\label{tab3}
		\centering
		\begin{tabular}{ccccc}
			\hline
			 & $s_1$ & $s_2$ & $s_3$ & $s_4$\\
			\hline
			$v_1$ & (2560.97, 2800, 3484.58) & (618.57, 700, 884.27) & (3406.72, 3500, 3828.08) & (4812.15, 5250, 5974.03) \\
			$v_2$ & (13918.81, 14000, 17808.50) & (3323.98, 3500, 3835.46) & (16819.00, 17500, 20539.04) & (25343.82, 26250, 28794.55) \\
			$v_3$ & (1144.78, 1200, 1450.83) & (298.19, 300, 303.49) & (1323.77, 1500, 1683.88) & (2089.08, 2250, 2479.93) \\
			$v_4$ & (701.05, 800, 1015.21) & (176.64, 200, 258.50) & (973.31, 1000, 1174.42) & (1427.83, 1500, 1936.48) \\
			$v_5$ & (1717.10, 2000, 2297.11) & (482.74, 500, 590.05) & (2468.10, 2500, 2889.79) & (3211.60, 3750, 4865.82) \\
			$v_6$ & (210.18, 240, 274.81) & (51.30, 60, 77.00) & (283.40, 300, 375.52) & (403.55, 450, 568.21) \\
			$v_7$ & (399.99, 400, 413.68) & (96.64, 100, 107.72) & (483.87, 500, 576.55) & (689.23, 750, 881.02) \\
			$v_8$ & (7849.11, 8000, 10081.39) & (1817.86, 2000, 2204.68) & (9113.32, 10000, 12923.71) & (14511.28, 15000, 16312.29) \\
			\hline
		\end{tabular}
	\end{table*}
 
 	Fig. \ref{fig2}(c) depicts the workflow scheduling results based on the random scheduling strategy \cite{ZHOU20206154}. It randomly schedules each task to their corresponding severs, and all tasks are executed according to their data dependencies. The fuzzy completion time ${\tilde t_{t}}$ of the workflow application is (11599.95\textit{s}, 11599.96\textit{s}, 11613.64\textit{s}), which exceeds the corresponding deadline (\textit{i.e., $7.2 \times {10^3}$s}). The fuzzy execution cost ${\tilde c_{t}}$ based on the random scheduling strategy is (38.61\textit{\$}, 38.84\textit{\$}, 39.6\textit{\$}), whose equivalent defuzzified value is 39.14\textit{\$}. 
 	Fig. \ref{fig2}(d) depicts the optimal workflow scheduling. The fuzzy completion time ${\tilde t_{t}}$ is (7036.56\textit{s}, 7052.69\textit{s}, 7129.24\textit{s}), which meets its deadline constraint. The fuzzy execution cost ${\tilde c_{t}}$  is (28.41\textit{\$}, 29.84\textit{\$}, 32.79\textit{\$}), whose equivalent  defuzzified value  is 30.93\textit{\$}. The optimal execution cost is significantly reduced by 21\% compared to that based on the random scheduling strategy.

	\section{Our Proposed Intelligent Decision Making Systems}
	
	The goal of a workflow scheduling strategy, $\Psi  = \left( {\bm{W},\bm{S},\bm{M},{\tilde t_{t}},{\tilde c_{t}}} \right)$, is to find the best mapping $\bm{M}$ from all tasks in a workflow $\bm{W}$ to different servers in edge-cloud environments $\bm{S}$, where the workflow execution cost ${\tilde c_{t}}$ is optimal within its corresponding deadline ${\tilde t_{t}}$. The tasks on a server have their strict execution order based on the data dependencies.
	A task can be executed on different servers, and a server also can process many tasks. Therefore, finding the best mapping from all tasks to different servers is a NP-hard problem \cite{HOSSEINISHIRVANI2020103501}.
	PSO is one of the effective algorithms to address such problems. Therefore, we propose a workflow scheduling strategy based on the modified PSO algorithm (\textit{i.e.}, ADPSO). The traditional PSO algorithm is introduced first, followed by a detailed description of ADPSO.

	\subsection{Traditional PSO Algorithm}
	
	PSO is an efficient evolutionary technique inspired by the social behavior of bird flocks. Kennedy and Eberhart first presented the PSO algorithm in 1995 \cite{488968}, which has been broadly investigated and utilized ever since. The particle is the most significant concept in PSO, which usually represents a candidate solution for an optimization problem. Each particle $Q_i^t=(X_i^t,V_i^t)$ in a population at the $t^{\rm{th}}$ iteration has its own position $X_i^t=(x_{i1}^t,x_{i2}^t, \ldots ,x_{iy}^t)$ and velocity $V_i^t=(v_{i1}^t,v_{i2}^t, \ldots ,v_{iz}^t)$, which will determine their direction and magnitude at the next iteration. The velocity of each particle is affected by their personal best particle $pB_i^t$ and the global best particle $gB^t$. Each particle constantly updates their own velocity and position in the potential solution space to obtain better fitness. The iterative update of velocities and positions for each particle are determined by Eqs. (\ref{eq24}) and (\ref{eq25}), respectively.
		   
	\begin{gather}
	V_i^{t + 1} = w \cdot V_i^t + {c_1}{r_1}({pB_i^t - X_i^t}) + {c_2}{r_2}({gB^t - X_i^t}), \label{eq24} \\
	X_i^{t + 1} = X_i^t + V_i^{t + 1}, \label{eq25}
	\end{gather} 
	
	\noindent where $w$ is an inertia weight, which determines to what extent the velocity of the current particles will affect the corresponding particles of the next generation, having a great impact on the convergence of PSO; $c_1$ and $c_2$ are acceleration coefficients, which denote the cognitive ability of a particle for its personal and global best particle, respectively; $r_1$ and $r_2$ are the random numbers on the interval [0,1), used to enhance the searching ability of PSO.
	
	The traditional PSO algorithm is designed for continuous optimization problems. However, workflow scheduling in edge-cloud environments is a discrete optimization problem. Therefore, an applicable PSO-based algorithm with new problem encoding and population update needs to be further adjusted.

	\subsection{ADPSO}
	
	The proposed ADPSO are described from five aspects: problem encoding, fitness function, population update, mapping from a particle to a fuzzy scheduling, and parameter settings.
	
	\subsubsection{Problem encoding}
	
	Problem encoding affects the searchability of  a PSO-based algorithm, which is expected to meet three major principles: \textit{Viability}, \textit{Completeness}, and \textit{Non-redundancy} \cite{6998950}. Inspired by the work in \cite{6782394}, an order-server nesting strategy is developed to encode the cost-driven workflow scheduling in uncertain edge-cloud environments. In particular, the $i^{\rm{th}}$ particle in the $t^{\rm{th}}$ iteration (\textit{i.e.}, $P_i^t$) is denoted by Eq. (\ref{eq26}).
	
	\begin{equation}\label{eq26}
	P_i^t = \left( {{{\left( {{\chi _{i1}},{s_{i1}}} \right)}^t},{{\left( {{\chi _{i2}},{s_{i2}}} \right)}^t}, \ldots ,{{\left( {{\chi _{i\left| \bm{V} \right|}},{s_{i\left| \bm{V} \right|}}} \right)}^t}} \right),
	\end{equation} 
	
	\noindent where ${\left( {{\chi _{ij}},{s_{ij}}} \right)^t},j = 1,2,...,|\bm{V}|$, indicates the assignment of the task $v_j$, meaning that $v_j$ is executed on the server $s_{ij}$ with a specified order ${{\chi _{ij}}}$. There are two criteria for the task execution on a server as follows:  
	
	\textbf{Criterion 1:} If two concurrent tasks without data dependencies (\textit{i.e.}, there are no direct or indirect  data dependencies between the tasks) are scheduled to the same server, the task with a larger order value will be processed earlier. 
    If two tasks have the same order value, the one entering pending queue earlier will be processed first. 
	
	\textbf{Criterion 2:} If two tasks with data dependencies are scheduled to the same server, the predecessor one is processed first.
	
	\begin{figure}[htb]
		\centering
		\includegraphics[scale=0.9]{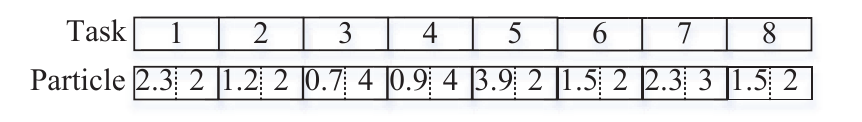}
		\caption{A encoded particle corresponding to the scheduling result of Fig. 2(d).}
		\label{fig3}
	\end{figure}
			
	Fig. \ref{fig3} depicts an encoded particle corresponding to the scheduling result of Fig. \ref{fig2}(d). After the task $v_1$ is executed, the tasks $v_5$ and $v_8$ are both scheduled to the server $s_2$. Since there are no data dependencies between $v_5$ and $v_8$, and $v_5$ has a larger order value (\textit{i.e.}, 3.9), it is processed first based on \textbf{Criterion 1}. The tasks $v_2$, $v_3$ and $v_6$ are next scheduled to the corresponding servers. At this moment, $v_8$, $v_2$ and $v_6$ are in the pending queue of the server $s_2$. Whilst $v_8$ has the same order value (\textit{i.e.}, 1.5) as $v_6$, it will be processed before $v_6$ because it enters the pending queue of $s_2$ earlier. The remaining tasks are similarly processed based on the two criteria.

	\subsubsection{Fitness function}
	
	Fitness function is used to evaluate the performance of particles. In general, a particle with a small fitness value represents a better candidate solution. This study aims to minimize the fuzzy total execution cost ${\tilde c_{t}}$ of scheduling a workflow within its deadline $D(\bm{W})$. Therefore, a particle corresponding to a scheduling result with a smaller fuzzy execution cost ${\tilde c_{t}}$ can be regarded as a better solution. However, the problem encoding strategy may not meet the \textit{Viability} principle, which dictates that the fuzzy completion time ${\tilde t_{t}}$ of a workflow must not exceed its deadline $D\left(\bm{W} \right)$.    Therefore, we compare the performance of two particles following three different situations.
		
	\textbf{Situation 1}: Both particles corresponding to the scheduling results  are feasible. The one with a smaller fuzzy total execution cost ${\tilde c_{t}}$ is deemed better, and the fitness function is defined by Eq. (\ref{eq27}).
	
	\begin{equation}\label{eq27}
	F\left( {{P_i}} \right) = {\tilde c_{t}}\left( {{P_i}} \right) \Rightarrow m\left( {{{\tilde c}_{t}}\left( {{P_i}} \right)} \right) + \eta  \cdot std\left( {{{\tilde c}_{t}}\left( {{P_i}} \right)} \right).
	\end{equation} 
	
	\textbf{Situation 2}: One particle corresponding to the scheduling result  is feasible, and the other is infeasible. The feasible particle is naturally deemed better, and the fitness function is defined by Eq. (\ref{eq28}).
	
	\begin{equation}\label{eq28}
	\begin{array}{l}
	F\left( {P_i^t} \right) = \left\{ {\begin{array}{*{20}{c}}
		{{{\tilde c}_t}\left( {P_i^t} \right),{{\tilde t}_t}\left( {P_i^t} \right) \le D\left( w \right)} \vspace{1ex}\\
		{{\kern 1pt} {\kern 1pt} {\kern 1pt} {\kern 1pt} {\kern 1pt} {\kern 1pt} {\kern 1pt} {\kern 1pt} {\kern 1pt} {\kern 1pt} \infty {\kern 1pt} {\kern 1pt} {\kern 1pt} {\kern 1pt} {\kern 1pt} {\kern 1pt} {\kern 1pt} {\kern 1pt} {\kern 1pt} {\kern 1pt} {\kern 1pt} ,{{\tilde t}_t}\left( {P_i^t} \right) > D\left( w \right)}
		\end{array}} \right.\vspace{1ex}\\
	\Rightarrow \left\{ {\begin{array}{*{20}{c}}
		{m\left( {{{\tilde c}_t}\left( {P_i^t} \right)} \right) + \eta  \cdot std\left( {{{\tilde c}_t}\left( {P_i^t} \right)} \right),{t^u}\left( {P_i^t} \right) \le D\left( w \right)} \vspace{1ex};\\
		{{\kern 1pt} {\kern 1pt} {\kern 1pt} {\kern 1pt} {\kern 1pt} {\kern 1pt} {\kern 1pt} {\kern 1pt} {\kern 1pt} {\kern 1pt} {\kern 1pt} {\kern 1pt} {\kern 1pt} {\kern 1pt} {\kern 1pt} {\kern 1pt} {\kern 1pt} {\kern 1pt} {\kern 1pt} {\kern 1pt} {\kern 1pt} {\kern 1pt} {\kern 1pt} {\kern 1pt} {\kern 1pt} {\kern 1pt} {\kern 1pt} {\kern 1pt} {\kern 1pt} {\kern 1pt} {\kern 1pt} {\kern 1pt} {\kern 1pt} {\kern 1pt} {\kern 1pt} {\kern 1pt} {\kern 1pt} {\kern 1pt} {\kern 1pt} {\kern 1pt} {\kern 1pt} {\kern 1pt} {\kern 1pt} {\kern 1pt} {\kern 1pt} {\kern 1pt} {\kern 1pt} {\kern 1pt} {\kern 1pt} {\kern 1pt} {\kern 1pt} {\kern 1pt} {\kern 1pt} {\kern 1pt} {\kern 1pt} {\kern 1pt} {\kern 1pt} \infty {\kern 1pt} {\kern 1pt} {\kern 1pt} {\kern 1pt} {\kern 1pt} {\kern 1pt} {\kern 1pt} {\kern 1pt} {\kern 1pt} {\kern 1pt} {\kern 1pt} {\kern 1pt} {\kern 1pt} {\kern 1pt} {\kern 1pt} {\kern 1pt} {\kern 1pt} {\kern 1pt} {\kern 1pt} {\kern 1pt} {\kern 1pt} {\kern 1pt} {\kern 1pt} {\kern 1pt} {\kern 1pt} {\kern 1pt} {\kern 1pt} {\kern 1pt} {\kern 1pt} {\kern 1pt} {\kern 1pt} {\kern 1pt} {\kern 1pt} {\kern 1pt} {\kern 1pt} {\kern 1pt} {\kern 1pt} {\kern 1pt} {\kern 1pt} {\kern 1pt} {\kern 1pt} {\kern 1pt} {\kern 1pt} {\kern 1pt} {\kern 1pt} {\kern 1pt} {\kern 1pt} {\kern 1pt} {\kern 1pt} {\kern 1pt} {\kern 1pt} {\kern 1pt} {\kern 1pt} {\kern 1pt} {\kern 1pt} {\kern 1pt} {\kern 1pt} {\kern 1pt} ,{t^u}\left( {P_i^t} \right) > D\left( w \right)}.
		\end{array}} \right.
	\end{array}
	\end{equation} 
	
	\textbf{Situation 3}:  Both particles corresponding to the scheduling results  are infeasible. The one with less fuzzy completion time ${\tilde t_{t}}$ is deemed better, which is more likely to become feasible after update operations. The fitness function is defined by Eq. (\ref{eq29}).
	
	\begin{equation}\label{eq29}
	F\left( {{P_i}} \right) = {\tilde t_{t
	}}\left( {{P_i}} \right) \Rightarrow {t^u}\left( {{P_i}} \right).
	\end{equation} 
	
	\subsubsection{Population update}
	
	The update of each particle is affected by  three factors: \textit{inertia}, \textit{individual cognition}, and \textit{social cognition} \cite{699146}. 
	To strengthen the searchability and avoid premature convergence of the proposed scheduling strategy, ADPSO employs the mutation operator and crossover operator of GA. The iterative update of the $i^{\rm{th}}$ particle at the ${(t+1)}^{\rm{th}}$ iteration for the workflow scheduling is defined as Eq. (\ref{eq30}).
	
	\begin{equation}\label{eq30}
	P_i^{t + 1} = o{p^{cr}}({o{p^{cr}}({o{p^{mu}}({P_i^t,w,r}),pB_i^t,{c_1},{r_1}}),g{B^t},{c_2},{r_2}}),
	\end{equation} 
	
	\noindent where $ {op}^{mu}() $ and $ {op}^{cr}() $ are mutation operation and crossover operation, respectively; $w$ is an inertia weight; $c_1$ and $c_2$ are acceleration coefficients; $r$, $r_1$ and $r_2$ are random numbers generated from the interval $[0,1)$.
	
	For the \textit{inertia} part, the mutation operator of GA \cite{7274769} is  introduced to perform the updating as Eq. (\ref{eq31}).
	
	\begin{equation}\label{eq31}
	A_i^{t + 1} = o{p^{mu}}\left( {P_i^t,w,r} \right) = \left\{ \begin{array}{l}
	{M_u}\left( {P_i^t} \right),r < w \vspace{1ex};\\
	{\kern 1pt} {\kern 1pt} {\kern 1pt} {\kern 1pt} {\kern 1pt} {\kern 1pt} {\kern 1pt} {\kern 1pt} {\kern 1pt} {\kern 1pt} {\kern 1pt} P_i^t{\kern 1pt} {\kern 1pt} {\kern 1pt} {\kern 1pt} {\kern 1pt} {\kern 1pt} {\kern 1pt} {\kern 1pt} {\kern 1pt} {\kern 1pt} {\kern 1pt} {\kern 1pt} {\kern 1pt} {\kern 1pt} {\kern 1pt} ,otherwise.
	\end{array} \right.
	\end{equation}
	
	\noindent Where $M_u()$ denotes the dual mutation operator, 
	which includes the neighborhood mutation operator for the task order and the adaptive multi-point mutation operator for the number of servers. 
	
	\begin{figure}[htb]
		\centering
		\subfigure[Neighborhood mutation operator for the task order ]{\includegraphics[scale=0.9]{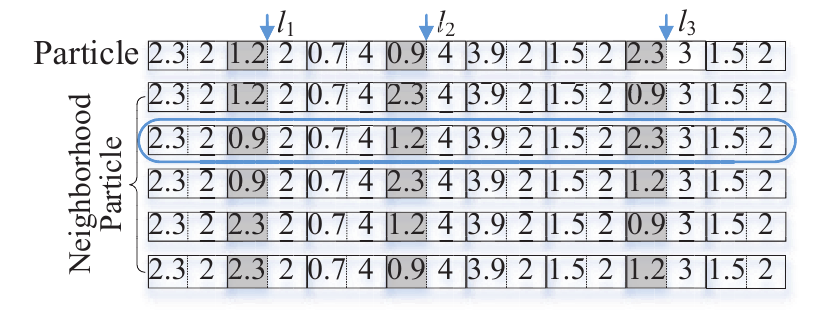}
			\label{fig4a}}
		\hfil
		\subfigure[Adaptive multi-point mutation operator for number of servers]{\includegraphics[scale=0.9]{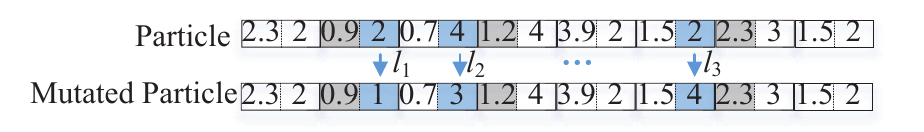}
			\label{fig4b}}
		\caption{Dual mutation operator.}
		\label{fig4}
	\end{figure}
		
	The neighborhood mutation operator randomly chooses three locations in a particle, and generates all sort combinations for the task order in the corresponding field. Then, it randomly selects a particle from the sort combinations as the generated one for feeding to the adaptive multi-point mutation operator.
	Fig. \ref{fig4a} depicts the neighborhood mutation operator. It randomly chooses the locations $l_1, l_2$ and $l_3$, and generates all sort combinations for the task order. It then randomly selects the second combination as the one for the adaptive multi-point mutation operator.
	
	The adaptive multi-point mutation operator randomly chooses $k$ locations (\textit{i.e.}, number of mutations) in a particle, and mutates each location's number of servers in the interval $[1,|\bm S|]$. Fig. \ref{fig4b} depicts this adaptive multi-point mutation operator. It randomly chooses $3$ locations (\textit{i.e.,} $l_1, l_2$ and $l_3$), and mutates the corresponding number of servers from (2,4,2) to (1,3,4).

	For \textit{individual cognition} and \textit{social cognition} parts, we introduce the crossover operator of GA to update the corresponding part of Eq. (\ref{eq24}). The updating process is determined by Eqs. (\ref{eq32}) and (\ref{eq33}).
	
	\begin{equation}\label{eq32}
	\begin{array}{cl}
		B_i^{t + 1} &= o{p^{cr}}({A_i^{t + 1},pB_i^t,{c_1},{r_1}})\vspace{1ex}\\
		 &= \left\{ \begin{array}{l}
		{C_r}({A_i^{t + 1},pB_i^t}),{r_1} < {c_1} \vspace{1ex};\\
		{\kern 1pt} {\kern 1pt} {\kern 1pt} {\kern 1pt} {\kern 1pt} {\kern 1pt} {\kern 1pt} {\kern 1pt} {\kern 1pt} {\kern 1pt} {\kern 1pt} {\kern 1pt} {\kern 1pt} {\kern 1pt} {\kern 1pt} {\kern 1pt} {\kern 1pt} {\kern 1pt} {\kern 1pt} {\kern 1pt} {\kern 1pt} {\kern 1pt} {\kern 1pt} {\kern 1pt} {\kern 1pt} {\kern 1pt} {\kern 1pt} A_i^{t + 1}{\kern 1pt} {\kern 1pt} {\kern 1pt} {\kern 1pt} {\kern 1pt} {\kern 1pt} {\kern 1pt} {\kern 1pt} {\kern 1pt} {\kern 1pt} {\kern 1pt} {\kern 1pt} {\kern 1pt} ,otherwise.
	\end{array} \right.
	\end{array}
	\end{equation}
	
	\begin{equation}\label{eq33}
	\begin{array}{cl}
		P_i^{t + 1} &= o{p^{cr}}({B_i^{t + 1},g{B^t},{c_2},{r_2}})\vspace{1ex}\\
		 &= \left\{
		\begin{array}{l}
		{C_r}({B_i^{t + 1},g{B^t}}),{r_2} < {c_2} \vspace{1ex};\\
		{\kern 1pt} {\kern 1pt} {\kern 1pt} {\kern 1pt} {\kern 1pt} {\kern 1pt} {\kern 1pt} {\kern 1pt} {\kern 1pt} {\kern 1pt} {\kern 1pt} {\kern 1pt} {\kern 1pt} {\kern 1pt} {\kern 1pt} {\kern 1pt} {\kern 1pt} {\kern 1pt} {\kern 1pt} {\kern 1pt} {\kern 1pt} {\kern 1pt} {\kern 1pt} {\kern 1pt} B_i^{t + 1}{\kern 1pt} {\kern 1pt} {\kern 1pt} {\kern 1pt} {\kern 1pt} {\kern 1pt} {\kern 1pt} {\kern 1pt} {\kern 1pt} {\kern 1pt} {\kern 1pt} {\kern 1pt} {\kern 1pt} {\kern 1pt} {\kern 1pt} {\kern 1pt} ,otherwise.
		\end{array} \right.
	\end{array}
	\end{equation}
	
	\noindent Where $C_r()$ is the two-point crossover operator. $C_r(A,B)$ randomly selects two locations in the particle A, and
	then replaces the corresponding segments between the two locations of A with the same interval in the particle B. Fig. 5 depicts
	this crossover operator in action. It randomly selects the locations $l_1$ and $l_2$
	in a mutated particle, and replaces the 
	segments between $l_1$ and $l_2$ with the same interval in
	$pB_i^t$ (or $gB^t$).

	\begin{figure}[htb]
		\centering
		\includegraphics[scale=0.9]{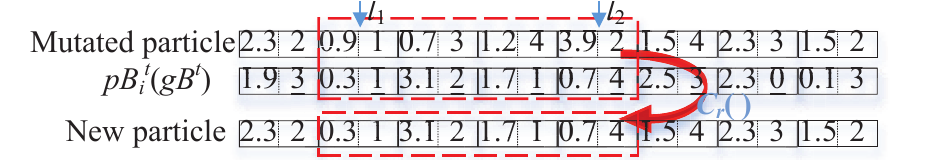}
		\caption{Crossover operator.}
		\label{fig5}
	\end{figure}
	
	\subsubsection{Mapping from a particle to a fuzzy scheduling}
	
	The mapping from a particle to a fuzzy scheduling in uncertain edge-cloud environments is summarized in Algorithm \ref{alg:one}. The inputs are the workflow $\bm W$, all available servers $\bm S$, and an encoded particle $P$. The output is the corresponding fuzzy workflow scheduling $\Psi =\left(\bm{W},\bm{S},\bm{M},{{{\tilde{t}}}_{t}},{{{\tilde{c}}}_{t}} \right)$ based on the particle $P$. It first initializes the mapping $\bm M$ to an empty set $null$ and ${{\tilde{c}}_{tran}}$ to (0,0,0). The fuzzy task processing time ${\tilde{t}_{com}}\left( \left| \bm{V} \right|\times \left| \bm{S} \right| \right)$ on different servers, and the fuzzy data transferring time ${\tilde{t}_{tran}}\left( \left| \bm{E} \right|,\left| \bm{S} \right|\times \left| \bm{S} \right| \right)$ between servers are calculated (line 3). 
	According to the encoded particle $P$, each task $v_i$ is scheduled on the server $s_j$ with the order $\chi_j$.	
	For a certain task $v_i$, its start time ${{\tilde{t}}_{start}}\left( {{v}_{i}} \right)$ is equal to the booting
	time ${\zeta}_j$ of the server $s_j$ if it is an entry task. Otherwise, the task cannot start until the last
	dataset is transferred to $s_j$ from its parents (line 13-22). Then, the end time ${{\tilde{t}}_{end}}\left( {{v}_{i}} \right)$ of $v_i$ is the sum of its start time ${{\tilde{t}}_{start}}\left( {{v}_{i}} \right)$ and  its processing time ${{\tilde{t}}_{com}}\left( {{v}_{i}} , {{s}_{j}} \right)$ on the server $s_j$ (line 24). 
	According to Eqs. (\ref{eq6}) and (\ref{eq7}), ${{{\tilde{t}}}_{t}}$ and ${{{\tilde{c}}}_{t}}$ are subsequently calculated. 
	Note that if the fuzzy completion time ${{{\tilde{t}}}_{t}}$ exceeds
	the corresponding deadline(\textit{i.e.}, $t^u > D(\bm W)$), the algorithm stops immediately
	and returns a symbolic value of False, meaning that this particle is infeasible
	(line 27-29). 
	Finally, it returns the fuzzy scheduling strategy $\Psi =\left(\bm{W},\bm{S},\bm{M},{{{\tilde{t}}}_{t}},{{{\tilde{c}}}_{t}} \right)$  if this particle is feasible (line 31).	
	
	\begin{algorithm}[t]
		\SetAlgoNoLine
		\KwIn{$\bm{W}, \bm{S}, P$.}
		\KwOut{$\Psi=\left(\bm{W},\bm{S},\bm{M},{{{\tilde{t}}}_{t}},{{{\tilde{c}}}_{t}} \right)$}
		
		$\textbf{Initialization}: \bm{M}\leftarrow null,{{\tilde{c}}_{tran}}\leftarrow (0,0,0)$.\\
		
		\Begin{
			Calculate ${{\tilde{t}}_{com}}\left( \left| \bm{V} \right|\times \left| \bm{S} \right| \right)$, ${{\tilde{t}}_{tran}}\left( \left| \bm{E} \right|,\left| \bm{S} \right|\times \left| \bm{S} \right| \right)$\;
			
			\For{i=$1$ to i=|$\bm{V}$|}{
				$\bm{M}=\bm{M}\cup \left( {{v}_{i}},{{s}_{j}} \right)$\;
				\eIf{$v_i$ is a entry task}{
					\If{$s_j$ is off}{
						Turn on $s_j$, ${\tilde{\zeta}}_{j}={\tilde{\varpi}}_{j}=(0,0,0)$\;
					}
					${{\tilde{t}}_{start}}\left( {{v}_{i}} \right)={\tilde{\zeta}} _{j}$\;
				}{
					$maxT=(0,0,0)$\;
					\ForEach{parent $v_p$ of $v_i$}{
						$maxT=maxT\vee \left( {{{\tilde{t}}}_{end}}\left( {{v}_{p}} \right)+{{{\tilde{t}}}_{tran}}\left( {{d}_{p,i}}, {{s}_{q}}, {{s}_{j}} \right) \right)$\;
						// $\left( {{v}_{p}},{{s}_{q}} \right),\left( {{v}_{i}},{{s}_{j}} \right)\in \bm{M}.$\\
						${{\tilde{c}}_{tran}}+=fuzzy\left( c_{q,j}^{tran}\cdot {{d}_{p,i}} \right)$\;
						// $fuzzy$(*) is the fuzzification function.\\
					}
					\If{$s_j$ is off}{
						Turn on $s_j$, ${\tilde{\zeta}}_{j}={\tilde{\varpi}}_{j}=maxT$\;
					}
					${{\tilde{t}}_{start}}\left( {{v}_{i}} \right)=maxT\vee{\tilde{\varpi}} _{j}$\;
				}
				${{\tilde{t}}_{end}}\left( {{v}_{i}} \right)={{\tilde{t}}_{start}}\left( {{v}_{i}} \right)+{{\tilde{t}}_{com}}\left( {{v}_{i}} , {{s}_{j}} \right)$\;
			}
			
			Calculate ${{{\tilde{t}}}_{t}},{{{\tilde{c}}}_{t}}$ based on (\ref{eq6}) and (\ref{eq7})\;
			
			\If{${{\tilde{t}}_{t}}>D(\bm W)$}{
				set $P$ as infeasible\;
				\Return False\;
			}
			\Return $\Psi =\left(\bm{W},\bm{S},\bm{M},{{{\tilde{t}}}_{t}},{{{\tilde{c}}}_{t}} \right)$\;
		}
		\caption{Mapping from a particle $P$ to a fuzzy scheduling strategy $\Psi$.}
		\label{alg:one}
	\end{algorithm}

	\subsubsection{Parameter settings}
	
	The inertia weight $w$ influences the convergence and searchability of PSO-based algorithm \cite{9380778}. A larger inertia weight helps the algorithm jumping out of local optima, improving its global searchability. By contrast, a smaller inertia weight improves the algorithm's local searchability. This study proposes a new adjustment mechanism that can adaptively adjust the value of inertia weight based on the particle's current state, thereby enhancing the algorithm's overall searchability, as shown in Eq. (\ref{eq34}).
	
	\begin{equation}\label{eq34}
		\left\{\begin{array}{c}
			w = {w_{max}} - ({w_{max}} - {w_{min}}) \times \exp (\frac{d(P_i^t)}{{d(P_i^t) - 1.01}}),\vspace{1ex}\\
			d(P_i^t) = \frac{div(gB^t, P_i^t)}{|P_i^t|},
		\end{array}\right.
	\end{equation}
	
	\noindent where ${w_{max}}$ and ${w_{min}}$ represent the predefined maximum and minimum values of $w$, $div(gB^t, P_i^t)$ represents the number of different encoding values between the current particle $P_i^t$ and the global best particle $gB^t$, and $|P_i^t|$ represents the size of the particle's encoding space. This mechanism can adaptively adjust the algorithm's searchability according to the difference between the global best particle and current particle. When $div(gB^t, P_i^t)$ is relatively small, it means that the difference between $gB^t$ and $P_i^t$ is small. Thus, the particle's local searchability is expected to be enhanced, increasing the algorithm's convergence. Conversely, a bigger value of $div(gB^t, P_i^t)$ means to increase the magnitude of $w$, enhancing the particle's global searchability.
	
	Regarding the adaptive multi-point mutation, the mutation number $k$ is adaptively adjusted according to the change of the inertia weight $w$, and its adjustment strategy is implemented by Eq. (\ref{eq35}).
	
	\begin{equation}\label{eq35}
	k = k_{max} + (k_{max} - k_{min}) \times \frac{w - w_{min}}{w_{max} - w_{min}},
	\end{equation}
	
	\noindent where $k_{max}$ and $k_{min}$ are the predefined maximum and minimum values of the mutation number $k$. The inertia weight $w$ has a positive effect on the mutation number $k$. When $w$ is large, the mutation number $k$ is increased to enhance the mutation ability so that the algorithm's global searchability can be intensified. On the contrary, if $w$ is small, then $k$ is decreased and as such, only limited mutation ability is reserved to maintain the diversity of the population.

	The acceleration coefficients $c_1,c_2$ are dynamically adjusted according to \cite{699146}, where $c_1^{{{s}}}$ and $c_2^{{{s}}}$ denote the start values of $c_1$ and $c_2$, and $c_1^{{{e}}}$ and $c_2^{{{e}}}$ denote their end values.

	\subsubsection{Algorithm flowcharts}
	
	\begin{figure}[htb]
		\centering
		\includegraphics[scale=0.9]{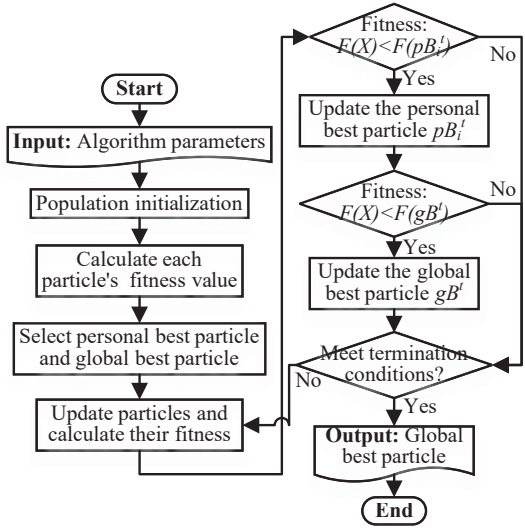}
		\caption{Flowchart of ADPSO algorithm.}
		\label{fig6}
	\end{figure}
	
	Fig. \ref{fig6} presents the flowchart of the ADPSO algorithm, which includes the following steps:
	
	\textbf{\textit{Step 1:}} Initialize the parameters of ADPSO, including the population size $\Gamma$, the maximum iteration number ${\Theta }$, inertia weight, and acceleration coefficients. Next, randomly generate the initial population.
	
	\textit{\textbf{Step 2:}} Calculate each particle's fitness value according to Eqs. (\ref{eq27}-\ref{eq29}). Each particle is set to its personal best particle and the particle with the smallest fitness value is set to the global best particle.
	
	\textit{\textbf{Step 3:}} Update all particles according to Eq. (\ref{eq30}), and recalculate the fitness value of each updated particle.
	
	\textit{\textbf{Step 4:}} Set the updated particle as the current personal best particle if its fitness value is less than the existing personal best; else, go to \textbf{\textit{Step 6}}.
	
	\textit{\textbf{Step 5:}} Set the updated particle as the global best particle if its fitness value is less than the existing global best.
	
	\textit{\textbf{Step 6:}} Check whether the termination condition is met; if so, output the global best particle and terminate, else, go back to \textbf{\textit{Step 3}}.

	\section{Performance Analysis}
	
	To validate the effectiveness of the workflow scheduling strategy based on the proposed ADPSO, experimental evaluations are carried out. In particular, the following Research Questions (\textbf{{RQs}}) are checked with the experiments conducted:
	
	\textbf{{RQ1:}} Compared with traditional PSO-based Algorithms, does ADPSO improve the searchability and convergence? ({Section \ref{sec5.1}})
	
	\textbf{{RQ2:}} In optimizing the fuzzy workflow execution cost, is ADPSO superior to other algorithms in terms of performance stability? ({Section \ref{sec5.2}})
	
	\textbf{{RQ3:}} In workflow scheduling with respect to given deadlines in uncertain edge-cloud environments, does ADPSO help reduce workflow execution cost? ({Section \ref{sec5.3}})
	

	\subsection{Basic experimental setup} \label{sec5.0}
	
	All experiments are run on the Win10 64-bit operating system with an Intel(R) Core(TM) i5-7200U CPU at 3.60 GHz and 16 GB RAM. Both ADPSO and all compared algorithms are implemented in Python 3.7. Parameters are set according to \cite{699146}, where $\Gamma$ = 100, ${\Theta }$ = 1000, $w_{max}$ = 0.9, $w_{min}$ = 0.4, $k_{max}$ = $|\bm{V}|/10$, $k_{min}$ = 1, $c_1^{s}$ = 0.9, $c_1^{e}$ = 0.2, $c_2^{s}$ = 0.4, and $c_2^{e}$ = 0.9.

	Five types of workflows are tested for this study, obtained from different scientific fields \cite{4723958}, including: CyberShake from earthquake science, Epigenomics from biogenetics, LIGO from gravitational physics, Montage from astronomy, and SIPHT from bioinformatics. Each type of workflow has different structures, numbers of tasks, and data transmissions between tasks, with detailed information stored in an XML file \cite{WorkflowHub}. 
	For each type of workflow, we choose three categories for the experiments: Tiny (approximately 30 tasks), Small (approximately 50 tasks) and Medium (approximately 100 tasks).
	
	There are three cloud servers ($s_1, s_2, s_3$) and two edge servers ($s_4, s_5$) in edge-could environments. Each server has specific processing ability and computation cost per time unit. We assume that $s_3$ has the most powerful processing ability, the processing time of the tasks on $s_3$ can be directly recorded from the corresponding XML file. Also, the processing capacity of $s_1$ or $s_2$ is approximately half or a quarter of that of $s_3$, while the processing capacity of $s_4$ or $s_5$ is about one-eighth or one-tenth of that of $s_3$. The computation cost per hour for $s_3$ is set to 15.5 \textit{\$}$\cdot$\textit{h}$^{-1}$, and the other severs' computation cost is approximately proportional to their processing abilities.

	The bandwidth and data transmission cost between different types of servers are set as Table \ref{tab1}.
	Each workflow $\bm W$ is assumed to have a corresponding deadline constraint in order to test the algorithm performance, set as Eq. (\ref{eq36}).
	
	\begin{equation}\label{eq36}
	D(\bm W) = 1.5 \times {H}(\bm W),
	\end{equation}
	
	\noindent where ${H}(\bm W)$ represents the execution time for scheduling $\bm W$ based on the HEFT algorithm \cite{993206}.
		
	{Section \ref{sec3.3}} has elaborated how to fuzzify the task processing time $t_{com}$ and data transferring time $t_{tran}$ to TFNs. The parameters $\delta_1$ and $\delta_2$ are empirically set to 0.85 and 1.2, respectively. For comparing the results effectively among different algorithms, conventional defuzzification method for TFNs \cite{ext-principle} is applied to the fuzzy execution cost and fuzzy completion time, where $\eta$ is set to 1. 
	
	\subsection{\textbf{RQ1.} Searchability and convergence} \label{sec5.1}
		
	\subsubsection{Compared algorithms} \label{sec5.1.2}
	
	In this subsection, we regard the traditional PSO \cite{488968} as the baseline algorithm for comparison, which adopts a similar order-server nesting encoding strategy. The server encoding value is computed continuously, and its rounding value is set as the server number. The update strategy of PSO is based on Eqs. (\ref{eq24}) and (\ref{eq25}), and the parameters $w, c_1, c_2$ are set according to \cite{699146}. By comparing the fuzzy execution cost of the candidate solutions and checking the deadline constraint's satisfaction for both algorithms, we analyze their performances in terms of the searchability and convergence.
	
	\subsubsection{Results and analysis} \label{sec5.1.3}
	
\begin{figure*}[!h]
		\centering
		\includegraphics[scale=0.38]{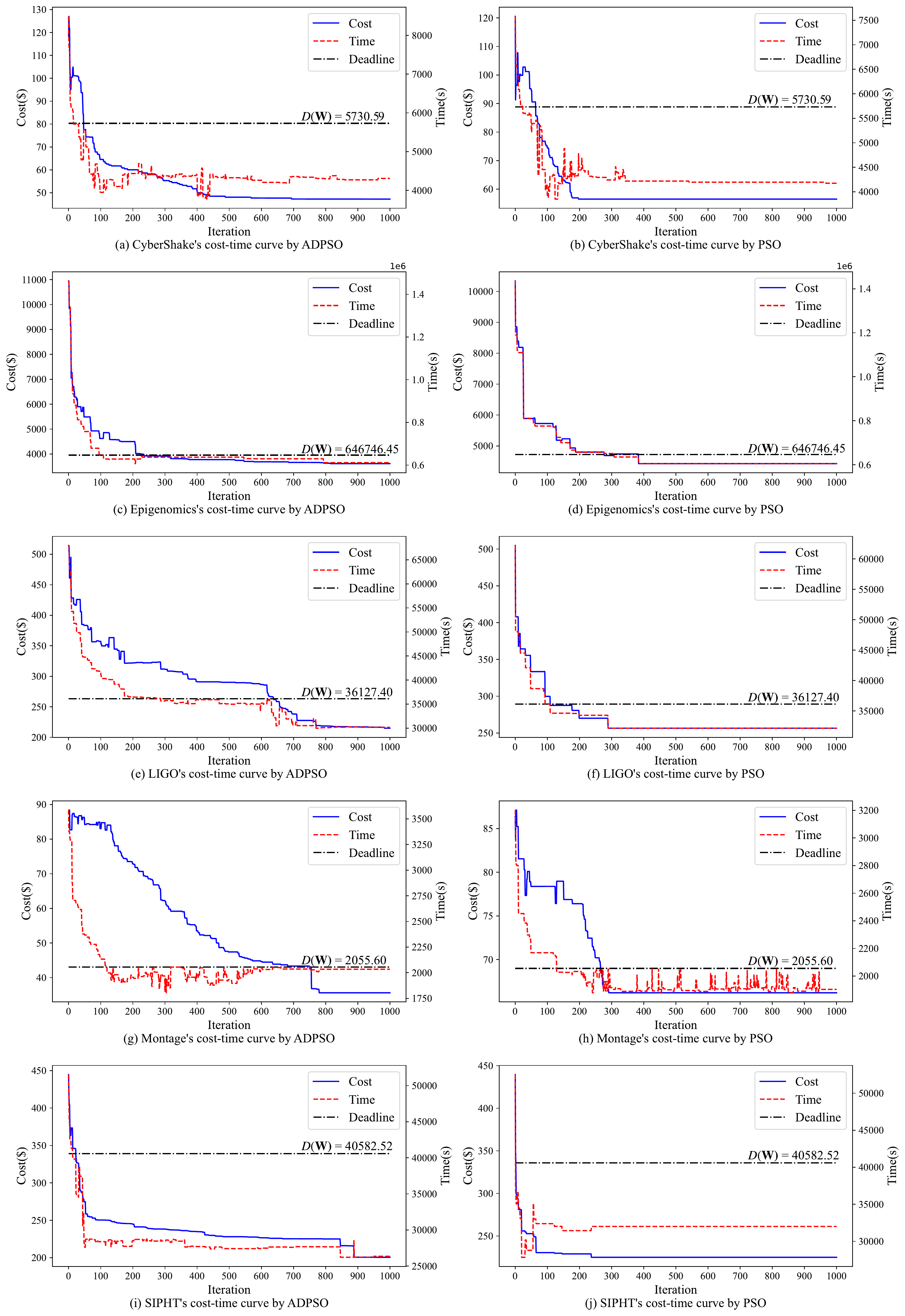}
		\caption{Time-cost iteration curves created by ADPSO and PSO for five types of medium workflow.}
		\label{fig7}
	\end{figure*}

    Different time units $\lambda_i$ are used depending on the size of the workflow. For the tiny and small workflows, the time unit $\lambda_i$ is one minute. For the medium workflow, the time unit $\lambda_i$ is one hour.
	We record both defuzzified fuzzy execution cost and fuzzy completion time of ADPSO and PSO algorithms over 1000 iterations. To clearly illustrate the convergence of fuzzy execution cost and to check whether fuzzy completion time meets the corresponding deadline constraint, Fig. \ref{fig7} shows the time-cost iteration curves created by ADPSO and PSO algorithms for five types of medium workflow, where a blue full-line,  a red dotted-line, and a horizontal black dotted-line indicate the workflow's execution cost, the completion time, and the corresponding deadline, respectively.
	
	ADPSO and PSO algorithms focus on decreasing the  workflow completion time during the early global searching, so that they could meet the deadline constraint and obtain a feasible solution. With the increasing of the iteration number, they become emphasizing on optimizing the execution cost within the given deadline while the local search becomes more accurate.	
	The number of iterations required for ADPSO and that for PSO to obtain the first feasible solution are similar. In particular, for CyberShake, as the execution cost is continuously optimizing, the completion time gradually stabilizes, fluctuating within the deadline.
	However, the scheduling strategy based on PSO becomes stable and converges to the global optimal particle at around the $200^{\rm {th}}$ iteration, whilst ADPSO can still update the global optimal particle even at the $700^{\rm {th}}$ iteration. 	
	In terms of the workflow execution cost, the scheduling strategy based on ADPSO is superior to that based on PSO.	
	Regarding LIGO, as the execution cost decreases, the workflow completion time overall tends to decrease as well. 
	Interestingly, the iteration number for convergence by PSO is less than that by ADPSO, but the convergence is reached prematurely as its performance on the execution cost is higher than that of ADPSO. 
	For Montage, from the perspective of fluctuation on the workflow completion time, although PSO may generate the global optimal particle at a latter stage, it has an inapparent improvement concerning the workflow execution cost.
	

	\subsection{\textbf{RQ2.} Stability on fuzzy execution cost} \label{sec5.2}
	
	\subsubsection{Compared algorithms}\label{sec5.2.2}
	
	Apart from the PSO algorithm introduced in {Section \ref{sec5.1.2}}, this subsection also describe the following two compared algorithms.
	(i)
	{Genetic Algorithm (GA) \cite{7274769}:} It employs the same order-server nesting encoding strategy as ADPSO. Following the classical update strategy of GA, the population is updated through binary tournament selection, two-point crossover operator, and mutation operator. Also, this algorithm takes the elitist preservation mechanism and completely copies the elite individuals into the next generation. The crossover probability and mutation probability of GA are set to 0.8 and 0.1, respectively \cite{7274769}.
	(ii)
	{Random Searching (RS) \cite{ZHOU20206154}:} RS also employs the same order-server nesting encoding strategy as ADPSO. The random searching strategy is adopted to generate the new population, with the order and server encoding of each particle being randomly created. In addition, there is no interference between iterations. The optimal solution in the population is produced at the end iteration.
	
	\subsubsection{Stability performance index}
	
	To test the stability of all algorithms, the sample variance vector $\bm{var}$ \cite{TYRISEVA20182187} is introduced to measure the performance of each algorithm. The sample variance $var_i$ of an algorithm is defined by Eq. (\ref{eq37}).
	
	\begin{equation}\label{eq37}
	var_i = \frac{\sum_{j=1}^h (F_j-\bar F)^2}{h-1},
	\end{equation}
	
	\noindent where $i=1, ..., 4$ corresponds to the four algorithms: ADPSO, PSO, GA, and RS. $h$ denotes the number of tests, $F_j$ denotes the fitness value at the $j^{\rm{th}}$ test, and $\bar F$ denotes the mean fitness value after $h$ tests.
	
	Due to the uncertainty of fuzzy workflow scheduling, the weak fluctuation can have a significant impact on the fuzzy execution cost. Hence, we consider the relative magnitudes of the variances among the four algorithms and normalize the sample variance vector $\bm{var}$ as follows:
	
	\begin{equation}\label{eq38}
	\bm{var'} = \frac{\bm{var}}{\Vert\bm{var}\Vert_2},
	\end{equation}
	
	\noindent where $\bm{var'}$ is the standardized sample variance vector, and $\Vert\cdot\Vert_2$ is the 2-norm. In general, the smaller the normalized sample variance is, the better stability the algorithm has.

	\subsubsection{Results and analysis}
	
	For each algorithm, ten independent repeated experiments are carried out, \textit{i.e.}, $h=10$. Table \ref{tab4} shows the normalized sample variances of the fuzzy execution cost for different types of workflow.
	To visually compare the normalized sample variances for each workflow among different algorithms, we annotate the minimum and maximum values of the normalized sample variances with the bold and underline fonts, respectively. 

	\begin{table}[htb]
		\renewcommand{\arraystretch}{1.3}
		\caption{Normalized sample variances of fuzzy execution cost for different types of workflow}
		\label{tab4}
		\centering
		\begin{tabular}{cccccc}
			\hline
			\multirow{2}{*}{Size} & \multirow{2}{*}{Workflows} & \multicolumn{4}{c}{Algorithms} \\
			\cline{3-6}
			& & ADPSO & PSO & GA & RS \\
			\hline
			\multirow{5}{*}{Tiny} & CyberShake & \uline{0.67} & \textbf{0.35} & 0.42 & 0.49 \\
			& Epigenomics & 0.61 & \uline{0.64} & \textbf{0.1} & 0.45 \\
			& LIGO & \textbf{0.07} & 0.1 & 0.08 & \uline{0.99} \\
			& Montage & \textbf{0.04} & 0.4 & 0.38 & \uline{0.83} \\
			& SIPHT & \textbf{0.02} & 0.1 & \uline{0.97} & 0.23 \\
			\hline
			\multirow{5}{*}{Small} & CyberShake & 0.57 & \textbf{0.35} & \uline{0.58} & 0.47 \\
			& Epigenomics & \textbf{0.03} & 0.14 & 0.04 & \uline{0.99} \\
			& LIGO & \textbf{0.04} & 0.09 & \uline{0.91} & 0.41 \\
			& Montage & \textbf{0.03} & \uline{0.84} & 0.11 & 0.53 \\
			& SIPHT & \textbf{0.19} & 0.25 & \uline{0.89} & 0.33 \\
			\hline
			\multirow{5}{*}{Medium} & CyberShake & 0.5 & 0.58 & \textbf{0.17} & \uline{0.62} \\
			& Epigenomics & \textbf{0.02} & 0.06 & 0.27 & \uline{0.96} \\
			& LIGO & \textbf{0.02} & \textbf{0.02} & \uline{0.87} & 0.49 \\
			& Montage & \textbf{0.08} & 0.62 & \uline{0.77} & 0.14 \\
			& SIPHT & 0.35 & \textbf{0.16} & 0.53 & \uline{0.76} \\
			\hline
		\end{tabular}
	\end{table}

	As shown in Table \ref{tab4}, ADPSO frequently obtains the  minimum normalized sample variances, followed by PSO. By contrast, GA and RS frequently obtain the maximum normalized sample variances, while ADPSO has the maximum normalized sample variance only for the case of tiny CyberShake. 	
	Furthermore, among these compared algorithms, ADPSO has the smallest normalized sample variance of the fuzzy execution cost among 12 types of workflow. For the case involving tiny workflow, ADPSO obtains the smallest normalized sample variances on 3 workflows. Considering the cases involving small or medium workflows, ADPSO obtains the smallest normalized sample variances over four workflows. In summary, ADPSO has the optimal normalized sample variances, which means that it has the best stability compared with other classical algorithms.

	\subsection{\textbf{RQ3.} Fuzzy workflow execution cost} \label{sec5.3}
	
	\subsubsection{Compared algorithms}
	
	The experiments conducted in this subsection also employ those compared algorithms adopted in 	{Section \ref{sec5.2.2}}.
	
	\subsubsection{Results and analysis}
	
	\begin{table*}[!t]
		\renewcommand{\arraystretch}{1.3}
		\caption{Fuzzy workflow execution cost of different algorithms for tiny workflows}
		\label{tab5}
		\centering
		\begin{tabular}{cccc}
			\hline
			Workflows & Algorithms & Optimal execution cost  and its corresponding fitness (\textit{\$})  & Mean execution cost and its corresponding fitness (\textit{\$}) \\
			\hline
			\multirow{4}{*}{CyberShake} & ADPSO [our] & \textbf{(8.94,8.98,9.08), 9.02} & \textbf{(11.25,11.48,12.14), 11.74} \\
			& PSO \cite{488968} & (11.08,11.19,11.60), 11.36 & (13.42,13.73,14.66), 14.09 \\
			& GA \cite{7274769} & (9.86,9.93,10.22), 10.05 & (12.18,12.49,13.31), 12.81 \\
			& RS \cite{7274769} & (17.96,18.66,20.65), 19.43 & (21.07,21.96,24.42), 22.91 \\
			\hline
			\multirow{4}{*}{Epigenomics} & ADPSO & \textbf{(146.64,154.25,174.53), 162.05} & \textbf{(151.50,159.47,181.36), 167.92} \\
			& PSO & (150.71,157.72,173.19), 163.51 & (157.01,164.49,177.16), 169.03 \\
			& GA & (162.41,166.14,177.28), 170.49 & (165.85,171.71,181.42), 175.17 \\
			& RS & (168.93,174.57,186.42), 178.98 & (173.47,180.76,198.27), 187.40 \\
			\hline
			\multirow{4}{*}{LIGO} & ADPSO & \textbf{(63.14,64.08,66.81), 65.14} & \textbf{(63.77,65.81,70.11), 67.41} \\
			& PSO & (64.80,67.05,72.19), 68.98 & (67.28,69.06,73.61), 70.80 \\
			& GA & (64.59,66.27,70.48), 67.88 & (66.90,68.27,72.09), 69.75 \\
			& RS & (79.77,81.74,86.06), 83.36* & (87.39,89.05,93.78), 90.88 \\
			\hline
			\multirow{4}{*}{Montage} & ADPSO & \textbf{(3.89,3.99,4.17), 4.05} & \textbf{(4.11,4.24,4.48), 4.32} \\
			& PSO & (4.79,4.92,5.33), 5.08 & (5.44,5.64,6.16), 5.84 \\
			& GA & (4.79,5.00,5.32), 5.11 & (5.52,5.76,6.30), 5.96 \\
			& RS & (12.45,13.07,14.53), 13.62 & (13.76,14.49,16.28), 15.17 \\
			\hline
			\multirow{4}{*}{SIPHT} & ADPSO & (56.26,58.05,64.33), 60.54 & \textbf{(56.58,58.38,64.79), 60.93} \\
			& PSO & \textbf{(58.51,59.79, 61.88), 60.53} & (59.41,60.90,63.78), 61.95 \\
			& GA & (60.69,60.84,61.31), 61.02 & (62.21,63.41,65.96), 64.36 \\
			& RS & (68.62,69.60,72.03), 70.52 & (70.62,71.72,74.62), 72.84 \\
			\hline
		\end{tabular}
	\end{table*}

	To compare the performances of these different algorithms for fuzzy workflow execution cost in the given uncertain edge-cloud environments, 
	we carry out 10 sets of independent repeated experiments for each algorithm with respect to cases involving different types of workflows. For each algorithm with one type of workflow, we record the optimal execution cost and its corresponding fitness (Unit: \textit{\$}), as well as the mean execution cost and its corresponding fitness (Unit: \textit{\$}) in the 10 independent repeated experiments.
	
	Table \ref{tab5} shows the fuzzy workflow execution cost of different algorithms for the tiny workflows. All the optimal solutions are listed in bold, and all the infeasible solutions are noted by `*'. 		
	Except for the case of SIPHT, ADPSO always obtains the optimal solution, beating PSO, GA and RS (with RS having the worst performance, and returning infeasible solutions). 
	The superior performance of ADPSO is mainly owing to 
	the improved encoding strategy, and the particle update mechanisms adopting the crossover operator and mutation operator of GA. 
	Thus, better scheduling strategy is attained by ADPSO while avoiding premature convergence to local optima. 
	Note that as the solution space of workflow scheduling is generally exponential, the random search strategy of RS is ineffective. As such, given limited population size and iteration number, it is difficult for RS to find high-quality solutions, even returning infeasible solutions.
	
	For the small workflows, experimental results 
	show that the performance of ADPSO is still the best, obtaining all the optimal solutions. 	
	For CyberShake, the optimal value of ADPSO is 26.4\% better than that of PSO, 15.9\% better than that of GA, and 125.5\% better than that of RS.	
	In addition, the mean execution cost value of ADPSO beats PSO, GA, and RS by 19.9\% for Montage, 16.3\% for CyberShake, and 117.6\% for Montage, respectively. 
	Indeed, ADPSO outperforms the other algorithms while scheduling CyberShake and Montage which involve the computing-intensive tasks.
	The scheduling results for medium workflows are almost the same as those for small workflows, with ADPSO achieving optimal solutions for all such workflows 
	By contrast, RS suffers from the worse performance with the expansion of task sizes, becoming almost impossible to obtain a feasible solution. 
	In short, the scheduling strategy based on ADPSO is able to reduce fuzzy workflow execution cost while achieving better scheduling outcomes compared with other algorithms.
	
	
	\subsection{Engineering Applications} \label{sec5.4}
	
	\begin{figure}[htb]
		\centering
		\includegraphics[scale=0.25]{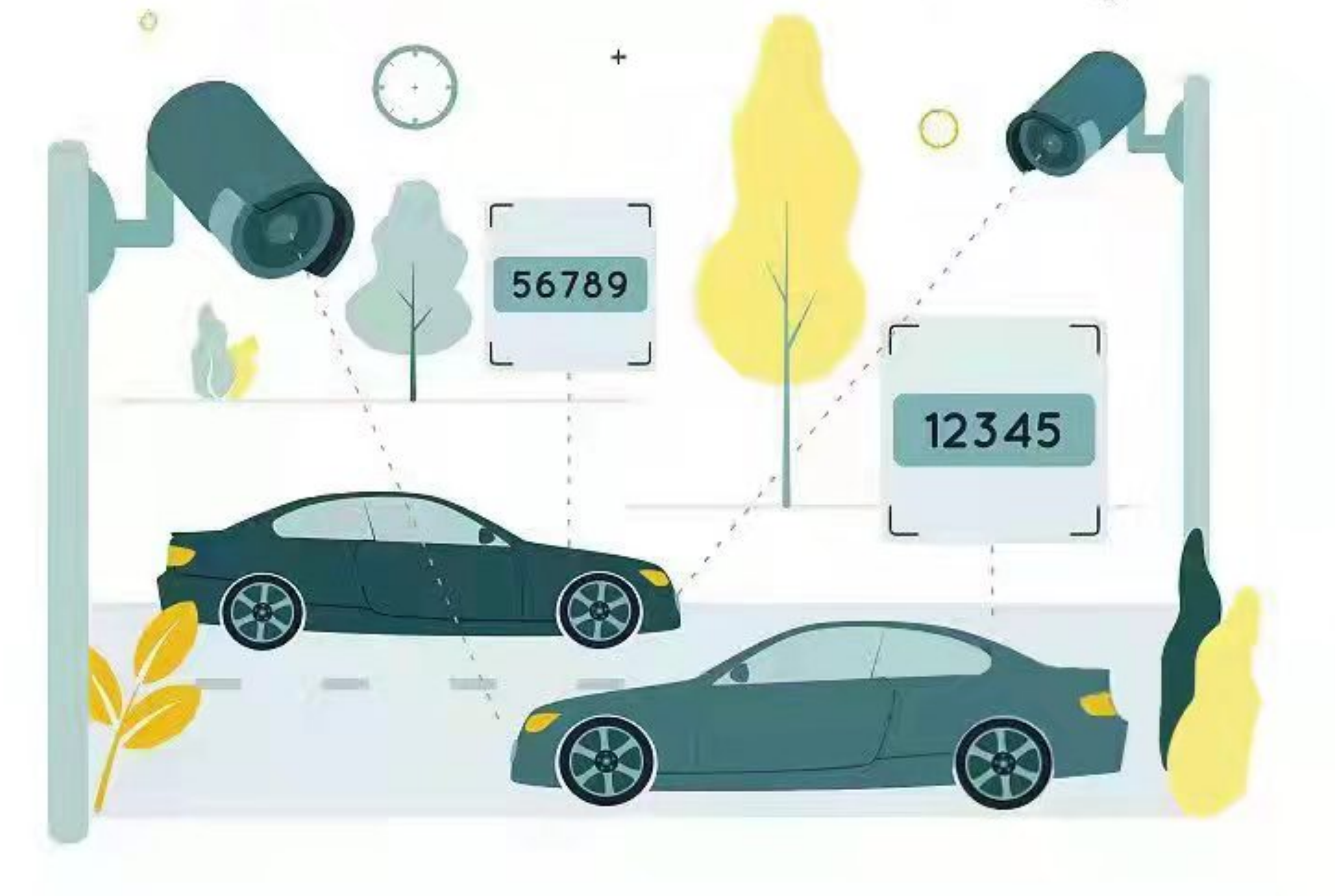}
		\caption{Vehicle identification application.}
		\label{fig8}
	\end{figure}
	
	Vehicle identification is one of the workflow applications in transportation systems, whose core technology is Deep Neural Networks (DNN) \cite{9416166,LIU2021107636}. 
	Traffic cameras with limited process capacity periodically
	record the images of on-road vehicles, and usually fail to complete the applications within their deadlines. Workflow decision making is one of the key issues to performance DNNs in vehicle identification applications. 
	Fig. \ref{fig8} presents the outline of the vehicle identification application. 
	
	\begin{figure}[htb]
		\centering
		\includegraphics[scale=0.04]{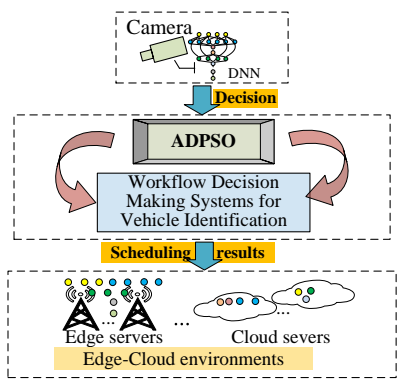}
		\caption{Scheduling strategy based on ADPSO for vehicle identification in uncertain edge-cloud environments.}
		\label{fig9}
	\end{figure}
	
	The uncertain environments have a great 
	impact on the system latency for such problems, which can easily lead to the misjudgement of the optimal scheduling.
	In addition, it is difficult to select an optimal layers-servers solution from the numerous combinations. Therefore, as shown in Fig. \ref{fig9}, we can employ the scheduling strategy based on ADPSO to make intelligent workflow decisions for vehicle identification applications, which can reduce the execution cost  mainly caused by layer computation and data transmission between layers within their deadlines, even in uncertain edge-cloud environments. 
	Complex DNN layers (tasks) in vehicle identification applications can be scheduled to the cloud for execution, while simple ones are processed on the edge. The cloud and edge  platforms collaborate with each other and execute the DNN layers with low system cost and latency.

	\section{Conclusion and the Future Work}
	
	This paper has proposed a cost-driven scheduling strategy based on  ADPSO for deadline-based workflow applications in uncertain edge-cloud environments. The work enables the reduction of workflow execution cost that is mainly caused by task computation and data transmission, while satisfying given deadlines.
	Experimental results have shown that the proposed strategy offers a better performance on scheduling computing-intensive workflows than the state-of-the-art approaches. Particularly, ADPSO can obtain the optimal fuzzy execution cost for almost all types of workflows investigated. 
	
	Our future work intends to improve the current strategy to address more types of workflow applications (than experimentally studied herein), as well as workflow ensembles in fuzzy edge-cloud environments. Also, to reflect the fact that in general, different tasks may have different requirements, we plan to refine the proposed approach to consider different load-to-cost ratios for different servers, while optimizing the overall workflow scheduling process.

	
	%

	

	%
	%

	\ifCLASSOPTIONcaptionsoff
	\newpage
	\fi

	
	
	%

	%
	\bibliography{T-SYSTEMS-21-11}
	
	\vspace{-30em}
	\begin{IEEEbiography}[{\includegraphics[width=1in,height=1.25in,clip,keepaspectratio]{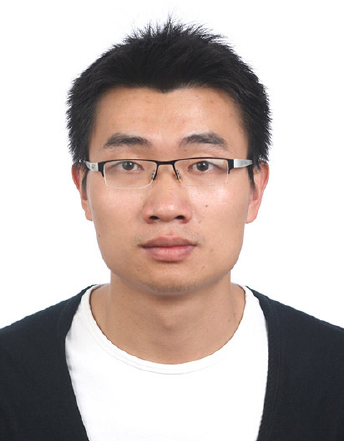}}]{Bing Lin}
		is currently an associate professor with the College of Physics and Energy at Fujian Normal University. Now he is a visiting scholar at School of Computer Science, Peking University. His research interest mainly includes parallel and distributed computing, computational intelligence, and data center resource management. He has published over thirty journals and conference articles, such as IEEE Transactions on Parallel and Distributed Systems, IEEE Transactions on Industrial Informatics, and IEEE Transactions on Network and Service Management.
	\end{IEEEbiography}
	
	\vspace{-30em}
	\begin{IEEEbiography}[{\includegraphics[width=1in,height=1.25in,clip,keepaspectratio]{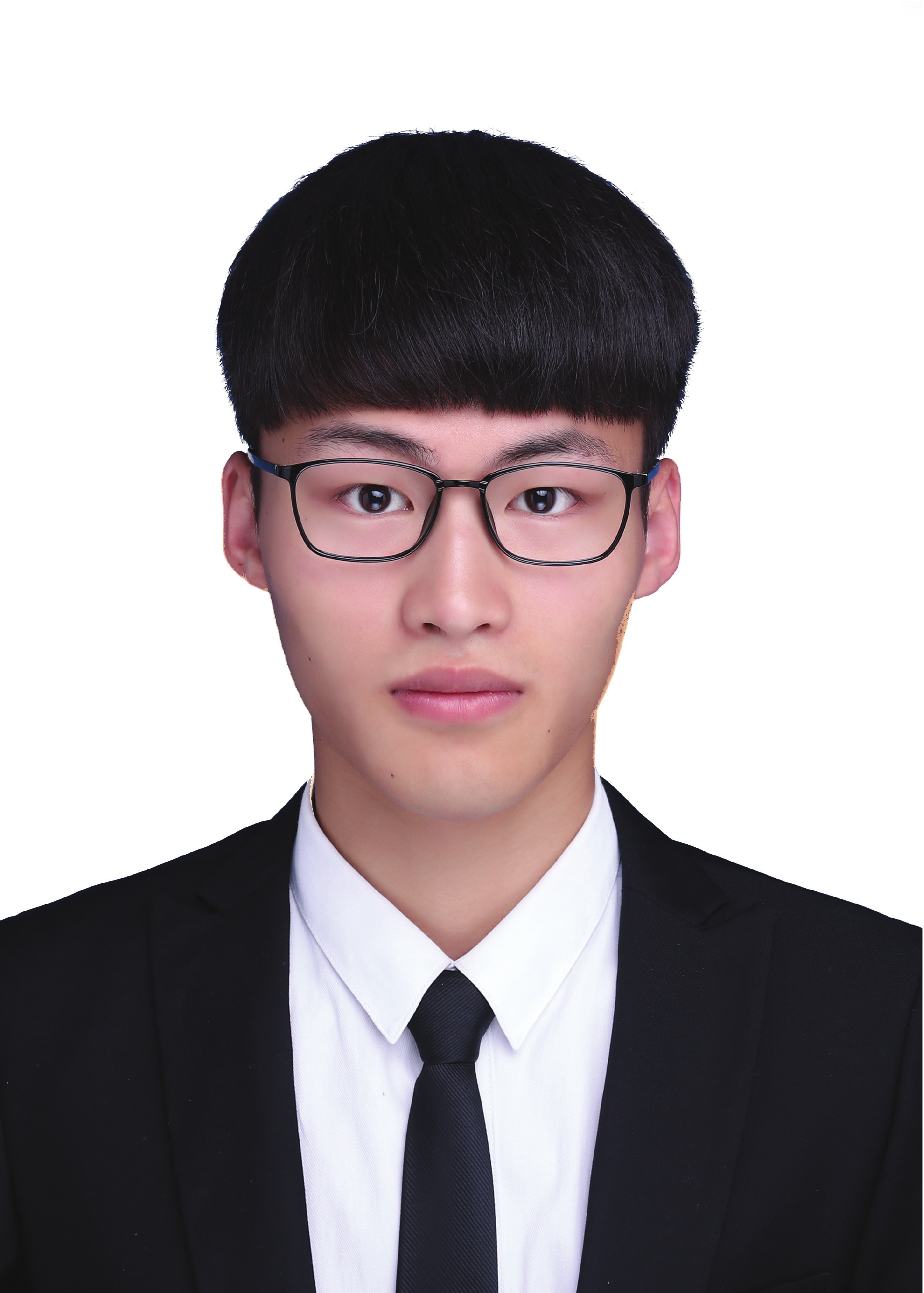}}]{Chaowei Lin}
		received the B.S. degree in Information and Computing Science from Fuzhou University, Fujian, China, in 2020. He is currently a postgraduate in Computer Software and Theory at the College of Mathematics and Computer Science, Fuzhou University, Fujian, China. His current research interests include workflow scheduling, computational intelligence, edge computing, and cloud computing.
	\end{IEEEbiography}

	\vspace{-30em}
	\begin{IEEEbiography}[{\includegraphics[width=1in,height=1.25in,clip,keepaspectratio]{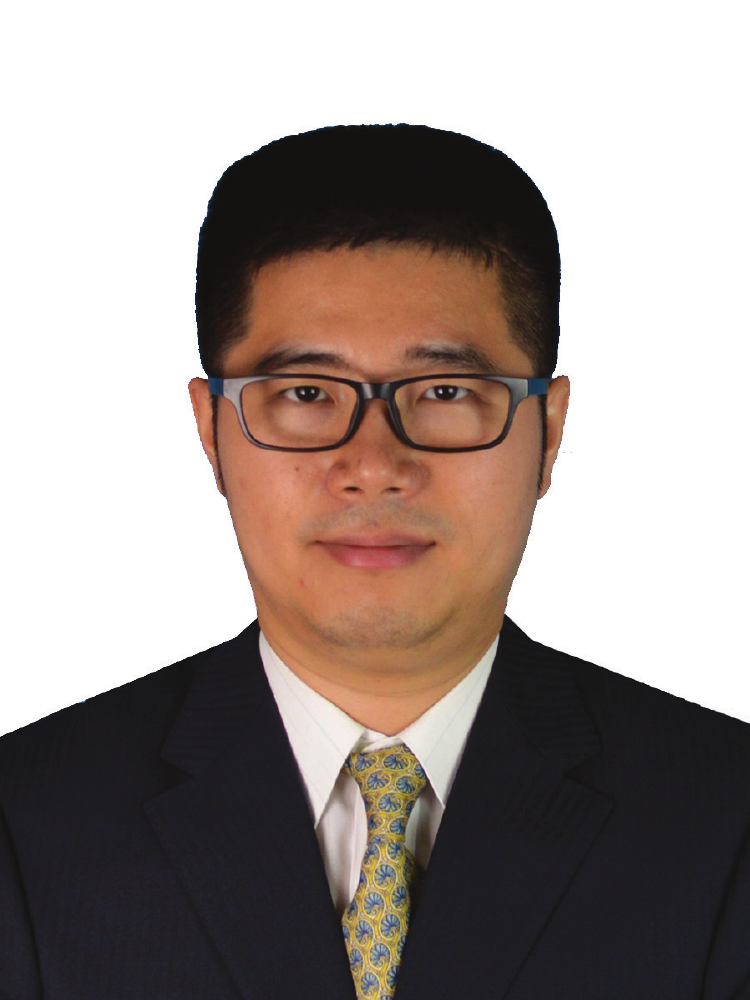}}]{Xing Chen}
		is a Professor at Fuzhou University, and the Director of Fujian Key Laboratory of Network Computing and Intelligent Information Processing. He received the B.S. degree and the Ph.D. degree from Peking University, in 2008 and 2013, respectively. He joined Fuzhou University since 2013.  He focuses on the software systems and engineering approaches for cloud and mobility. His current projects cover the topics from self-adaptive software, computation offloading, model driven approach and so on. He has published over 80 journal and conference articles, including IEEE Transactions on Parallel and Distributed Systems, IEEE Transactions on Cloud Computing, IEEE Transactions on Industrial Informatics, etc.
	\end{IEEEbiography}

\end{document}